\documentclass[aps,nofootinbib,longbibliography,notitlepage,superscriptaddress]{revtex4-1}

\usepackage[T1]{fontenc}
\usepackage[latin9]{inputenc}
\usepackage{color}
\usepackage{amsmath}
\usepackage{amssymb}
\usepackage{graphicx}
\usepackage{esint}
\usepackage{hyperref}
\allowdisplaybreaks

\makeatletter
\@ifundefined{textcolor}{}
{%
 \definecolor{BLACK}{gray}{0}
 \definecolor{WHITE}{gray}{1}
 \definecolor{RED}{rgb}{1,0,0}
 \definecolor{GREEN}{rgb}{0,1,0}
 \definecolor{BLUE}{rgb}{0,0,1}
 \definecolor{CYAN}{cmyk}{1,0,0,0}
 \definecolor{MAGENTA}{cmyk}{0,1,0,0}
 \definecolor{YELLOW}{cmyk}{0,0,1,0}
}

\usepackage{subfigure}

\makeatother

\begin{document}

\title{Stable and unstable cosmological models in bimetric massive gravity}

\author{Frank Koennig}
\email{koennig@thphys.uni-heidelberg.de}
\affiliation{Institut F\"ur Theoretische Physik, Ruprecht-Karls-Universit\"at
Heidelberg, \\
 Philosophenweg 16, 69120 Heidelberg, Germany}

\author{Yashar Akrami}
\email{yashar.akrami@astro.uio.no}
\affiliation{Institute of Theoretical Astrophysics, University of Oslo,
P.O. Box 1029 Blindern, N-0315 Oslo, Norway}

\author{Luca Amendola}
\email{l.amendola@thphys.uni-heidelberg.de}
\affiliation{Institut F\"ur Theoretische Physik, Ruprecht-Karls-Universit\"at
Heidelberg, \\
 Philosophenweg 16, 69120 Heidelberg, Germany}

\author{Mariele Motta}
\email{mariele.motta@unige.ch}
\affiliation{D\'epartement de Physique Th\'eorique and Center for Astroparticle Physics,
Universit\'e de Gen\`eve, Quai E. Ansermet 24, CH-1211 Gen\`eve 4, Switzerland}

\author{Adam R. Solomon}
\email{a.r.solomon@damtp.cam.ac.uk}
\affiliation{DAMTP, Centre for Mathematical Sciences, University of Cambridge,
\\
 Wilberforce Road, Cambridge CB3 0WA, United Kingdom}

\begin{abstract}
Nonlinear, ghost-free massive gravity has two tensor fields; when
both are dynamical, the mass of the graviton can lead to cosmic acceleration
that agrees with background data, even in the absence of a cosmological
constant. Here the question of the stability of linear perturbations
in this bimetric theory is examined. Instabilities are presented for several
classes of models, and simple criteria for the cosmological stability
of massive bigravity are derived. In this way, we identify a particular self-accelerating
bigravity model, infinite-branch bigravity (IBB), which exhibits both viable background evolution and
stable linear perturbations. We discuss the modified
gravity parameters for IBB, which do not reduce to the standard $\Lambda$CDM
result at early times, and compute the combined likelihood from measured
growth data and type Ia supernovae. IBB predicts a
present matter density $\Omega_{m0}=0.18$ and an equation of
state $w(z)=-0.79+0.21z/(1+z)$. The growth rate of structure is well-approximated at late times by $f(z)\approx\Omega_{m}^{0.47}[1+0.21z/(1+z)]$. The implications of the linear instability
for other bigravity models are discussed: the instability does not
necessarily rule these models out, but rather presents interesting
questions about how to extract observables from them when linear perturbation
theory does not hold. 
\end{abstract}
\maketitle

\section{Introduction}

Testing gravity beyond the limits of the Solar System is an important
task of present and future cosmology. The detection of any modification
of Einstein's gravity at large scales or in past epochs would be an extraordinary
revolution and change our view of the evolution of the Universe.

A theory of a massless spin-2 field is either described by general relativity \cite{Gupta:1954zz,Weinberg:1965rz,Deser:1969wk,Boulware:1974sr,Feynman:1996kb,2010GReGr..42..641D} or unimodular gravity \cite{2014PhRvD..89l4019B, 2014arXiv1406.7713B}.

Consequently, most modifications of gravity proposed so far introduce
one or more new dynamical fields, in addition to the massless metric
tensor of standard gravity. This new field is usually a scalar field,
typically through the so-called Horndeski Lagrangian \cite{Horndeski:1974,Deffayet:2011gz},
or a vector field, such as in Einstein-aether models (see Refs. \cite{Jacobson:2008aj,Solomon:2013iza}
and references therein). A complementary approach which has gained
significant attention in recent years is, rather than adding a new dynamical
field, to promote the massless spin-2 graviton of general relativity
to a massive one.

The history of massive gravity is an old one, dating back to 1939,
when the linear theory of Fierz and Pauli was published \cite{Fierz:1939ix}.
We refer the reader to the reviews \cite{2012RvMP...84..671H,deRham:2014zqa}
for a reconstruction of the steps leading to the modern approach,
which has resulted in a ghost-free, fully nonlinear theory of massive
gravity \cite{deRham:2010kj} (see also Refs. \cite{2010PhRvD..82d4020D,2010PhLB..693..334D,Hassan:2011vm,Hassan:2011hr,Maggiore:2014sia}).
A key element of these new forms of massive gravity is the introduction
of a second tensor field, or ``reference metric,'' in addition to
the standard metric describing the curvature of spacetime. When this
reference metric is fixed (e.g., Minkowski), this theory propagates
the five degrees of freedom of a ghost-free massive graviton.

However, the reference metric can also be made dynamical, as proposed
in Refs. \cite{2012JHEP...02..026H,2012JHEP...02..126H}. This promotes
massive gravity to a theory of \textit{bimetric} gravity. This theory
is still ghost free and has the advantage of allowing cosmologically viable
solutions. The cosmology of bimetric gravity has been studied in several
papers, e.g., in Refs. \cite{2012JCAP...03..042V,2013JHEP...03..099A,2013JCAP...10..046A,Comelli:2012db,2013arXiv1304.3920D,2012JHEP...03..067C,2012JHEP...01..035V}.
The main conclusion is that bimetric gravity allows for a cosmological
evolution that can approximate the $\Lambda$CDM universe and can
therefore be a candidate for dark energy without invoking a cosmological
constant. Crucially, the parameters and the potential structure leading
to the accelerated expansion are thought to be stable under quantum
corrections \cite{deRham:2013qqa}, in stark contrast to a cosmological
constant, which would need to be fine-tuned against the energy of the
vacuum \cite{Weinberg:1988cp,Martin:2012bt}.

Bimetric gravity has been successfully compared to background data
(cosmic microwave background, baryon acoustic oscillations, and type
Ia supernovae) in Refs. \cite{2012JCAP...03..042V,2013JHEP...03..099A},
and to linear perturbation data in Refs. \cite{Konnig:2014dna,2014arXiv1404.4061S}.
The comparison with linear perturbations has been undertaken on subhorizon scales assuming
a quasistatic (QS) approximation, in which the potentials are assumed
to be slowly varying. This assumption makes it feasible to derive
the modification to the Poisson equation and the anisotropic stress,
two functions of scale and time which completely determine observational
effects at the linear level.

The quasistatic equations are, however, a valid subhorizon approximation
only if the full system is stable for large wave numbers. Previous
work \cite{Comelli:2012db,DeFelice:2014nja,Comelli:2014bqa} has identified a region of instability in the past.\footnote{This should not be confused with the Higuchi ghost instability, which affects most massive gravity cosmologies and some in bigravity, but is, however, absent from the simplest bimetric models which produce $\Lambda$CDM-like backgrounds \cite{2013JCAP...12..002F}.} Here we investigate this problem in detail.
We reduce the linearized Einstein equations to two equations for the
scalar modes, and analytically determine the epochs of stability and
instability for all the models with up to two free parameters which
have been shown to produce viable cosmological background evolution.
The behavior of more complicated models can be reduced to these simpler
ones at early and late times.

We find that several models which yield sensible background cosmologies
in close agreement with the data are in fact plagued by an instability
that only turns off at recent times. This does not necessarily rule
these regions of the bimetric parameter space out, but rather presents
a question of how to interpret and test these models, as linear perturbation
theory is quickly invalidated. Remarkably, we find that only a particular
bimetric model --- the one in which only the $\beta_{1}$ and $\beta_{4}$
parameters are nonzero (that is, the linear interaction and the cosmological
constant for the reference metric are turned on) --- is stable and has a cosmologically viable background at
all times when the evolution is within a particular branch.  This shows
that a cosmologically viable bimetric model without an explicit cosmological constant
(by which we mean the constant term appearing in the Friedmann equation) does indeed exist, and raises the question of how to nonlinearly
probe the viability of other bimetric models.

This paper is part of a series dedicated to the cosmological perturbations
of bimetric gravity and their properties, following Ref. \cite{2014arXiv1404.4061S}.

\section{Background equations}

We start with the action of the form \cite{2012JHEP...02..126H} 
\begin{eqnarray}
S & = & -\dfrac{M_{g}^{2}}{2}\int d^{4}x\sqrt{-\det g}R(g)-\dfrac{M_{f}^{2}}{2}\int d^{4}x\sqrt{-\det f}R(f)\\
 & + & m^{2}M_{g}^{2}\int d^{4}x\sqrt{-\det g}\sum_{n=0}^{4}\beta_{n}e_{n}\left(\sqrt{g^{\alpha\beta}f_{\beta\gamma}}\right)+\int d^{4}x\sqrt{-\det g}\mathcal{L}_{m}(g,\Phi),
\end{eqnarray}
where $e_{n}$ are elementary symmetric polynomials and $\beta_{n}$
are free parameters. Here $g_{\mu\nu}$ is the standard metric coupled
to the matter fields $\Phi$ in the matter Lagrangian, $\mathcal{L}_{m}$,
while $f_{\mu\nu}$ is a new dynamical tensor field with metric properties.
In the following we express masses in units of $M_{g}$ and absorb
the mass parameter $m^{2}$ into the parameters $\beta_{n}$. The
graviton mass is generally of order $m^{2}\beta_{n}$. The action
then becomes 
\begin{eqnarray}
S & = & -\dfrac{1}{2}\int d^{4}x\sqrt{-\det g}R(g)-\dfrac{M_{f}^{2}}{2}\int d^{4}x\sqrt{-\det f}R(f)\\
 & + & \int d^{4}x\sqrt{-\det g}\sum_{n=0}^{4}\beta_{n}e_{n}\left(\sqrt{g^{\alpha\beta}f_{\beta\gamma}}\right)+\int d^{4}x\sqrt{-\det g}\mathcal{L}_{m}(g,\Phi).
\end{eqnarray}

There has been some discussion in the literature over how to correctly take square roots. We will find solutions in which $\det\sqrt{g^{-1}f}$ becomes zero at a finite point in time (and only at that time), and so it is important to determine whether to choose square roots to always be positive, or to change sign on either side of the $\det=0$ point. This was discussed in some detail in Ref. \cite{2013CQGra..30r4007G} (see also Ref. \cite{2014PhRvD..89b7502G}), where continuity of the vielbein corresponding to $\sqrt{g^{-1}f}$ demanded that the square root \textit{not} be positive definite. We will take a similar stance here, and make the only choice that renders the action differentiable at all times, i.e., such that the derivative of $\sqrt{g^{-1}f}$ with respect to $g_{\mu\nu}$ and $f_{\mu\nu}$ is continuous everywhere. In particular, using a cosmological background with $f_{\mu\nu} \equiv \operatorname{diag}(-X^2,b^2,b^2,b^2)$, this choice implies that we assume $\sqrt{-\det f}=Xb^3$, where $X=\dot{b}/\mathcal{H}$ with $\mathcal{H}$ is the $g$-metric Hubble rate. This is important because, as we will see later on, it turns out that in the cosmologically stable model, the $f$ metric bounces, so $X$ changes sign during cosmic evolution. Consequently the square roots will change sign as well, rather than develop cusps. Note that sufficiently small perturbations around the background will not lead to a different sign of this square root.

Varying the action with respect to $g_{\mu\nu}$, one obtains the
following equations of motion: 
\begin{equation}
R_{\mu\nu}-\dfrac{1}{2}g_{\mu\nu}R+\sum_{n=0}^{3}(-1)^{n}\beta_{n}g_{\mu\lambda}Y_{(n)\nu}^{\lambda}\left(\sqrt{g^{\alpha\beta}f_{\beta\gamma}}\right)=T_{\mu\nu}.\label{eq:eeg}
\end{equation}
Here the matrices $Y_{(n)\nu}^{\lambda}\left(\sqrt{g^{\alpha\beta}f_{\beta\gamma}}\right)$
are defined as, setting $\mathbb{X}=\left(\sqrt{g^{-1}f}\right)$,
\begin{align}
Y_{(0)}(\mathbb{X}) & =\mathbb{I},\\
Y_{(1)}(\mathbb{X}) & =\mathbb{X}-\mathbb{I}[\mathbb{X}],\\
Y_{(2)}(\mathbb{X}) & =\mathbb{X}^{2}-\mathbb{X}[\mathbb{X}]+\dfrac{1}{2}\mathbb{I}\left([\mathbb{X}]^{2}-[\mathbb{X}^{2}]\right),\\
Y_{(3)}(\mathbb{X}) & =\mathbb{X}^{3}-\mathbb{X}^{2}[\mathbb{X}]+\dfrac{1}{2}\mathbb{X}\left([\mathbb{X}]^{2}-[\mathbb{X}^{2}]\right)-\dfrac{1}{6}\mathbb{I}\left([\mathbb{X}]^{3}-3[\mathbb{X}][\mathbb{X}^{2}]+2[\mathbb{X}^{3}]\right),
\end{align}
where $\mathbb{I}$ is the identity matrix and $[...]$ is the trace
operator. Varying the action with respect to $f_{\mu\nu}$ we find
\begin{equation}
\bar{R}_{\mu\nu}-\dfrac{1}{2}f_{\mu\nu}\bar{R}+\dfrac{1}{M_{f}^{2}}\sum_{n=0}^{3}(-1)^{n}\beta_{4-n}f_{\mu\lambda}Y_{(n)\nu}^{\lambda}\left(\sqrt{f^{\alpha\beta}g_{\beta\gamma}}\right)=0,
\end{equation}
where the overbar indicates the curvature of the $f_{\mu\nu}$ metric.

The $f$-metric Planck mass, $M_{f}$, is a redundant parameter and
can be freely set to unity \cite{2012JCAP...12..021B}. To see this,
consider the rescaling $f_{\mu\nu}\rightarrow M_{f}^{-2}f_{\mu\nu}$.
The Ricci scalar transforms as $\bar{R}(f)\rightarrow M_{f}^{2}\bar{R}(f)$,
so the full Einstein-Hilbert term in the action becomes 
\begin{equation}
\frac{M_{f}^{2}}{2}\sqrt{-\det f}\bar{R}(f)\rightarrow\frac{1}{2}\sqrt{-\det f}\bar{R}(f).
\end{equation}
The other term in the action that depends on $f_{\mu\nu}$ is the
mass term, which transforms as 
\begin{equation}
\sum_{n=0}^{4}\beta_{n}e_{n}\left(\sqrt{g^{-1}f}\right)\rightarrow\sum_{n=0}^{4}\beta_{n}e_{n}\left(M_{f}^{-1}\sqrt{g^{-1}f}\right)=\sum_{n=0}^{4}\beta_{n}M_{f}^{-n}e_{n}\left(\sqrt{g^{-1}f}\right),
\end{equation}
where in the last equality we used the fact that the elementary symmetric
polynomials $e_{n}(\mathbb{X})$ are of order $\mathbb{X}^{n}$. Therefore,
by additionally redefining the interaction couplings as $\beta_{n}\rightarrow M_{f}^{n}\beta_{n}$,
we end up with the original bigravity action but with $M_{f}=1$ \footnote{Recall that we are expressing masses in units of the Planck mass,
$M_{g}$. In more general units, the redundant parameter is $M_{f}/M_{g}$.%
}.%
 Consequently we set $M_{f}=1$ in the following.

Let us now consider the background cosmology of bimetric gravity. We assume a spatially flat FLRW metric, 
\begin{equation}
ds_{g}^{2}=a^{2}(\tau)(-d\tau^{2}+dx_{i}dx^{i}),
\end{equation}
where $\tau$ is conformal time and an overdot represents the derivative
with respect to it. The second metric is chosen as 
\begin{equation}\label{metric_f}
ds_{f}^{2}=-\left[\dot{b}(\tau)^{2}/\mathcal{H}^{2}(\tau)\right]d\tau^{2}+b(\tau)^{2}dx_{i}dx^{i},
\end{equation}
where $\mathcal{H}\equiv\dot{a}/a$ is the conformal-time Hubble parameter
associated with the physical metric, $g_{\mu\nu}$. The particular
choice for the $f$-metric lapse, $f_{00}$, ensures that the Bianchi
identity is satisfied (see, e.g., Ref. \cite{2012JHEP...02..026H}).

Inserting the FLRW ansatz for $g_{\mu\nu}$ into Eq. (\ref{eq:eeg})
we get 
\begin{eqnarray}
3\mathcal{H}{}^{2} & = & a^{2}(\rho_{\mathrm{tot}}+\rho_{\mathrm{mg}}),\label{eq:fried-1}
\end{eqnarray}
where we define an effective massive-gravity energy density as 
\begin{equation}
\rho_{\mathrm{mg}}=B_{0}\equiv\beta_{0}+3\beta_{1}r+3\beta_{2}r^{2}+\beta_{3}r^{3}
\end{equation}
with 
\begin{equation}
r\equiv\frac{b}{a},
\end{equation}
while $\rho_{\mathrm{tot}}$ is the density of all other matter components
(e.g., dust and radiation). The total energy density follows the usual
conservation law, 
\begin{equation}
\dot{\rho}_{\mathrm{tot}}+3\mathcal{H}\rho_{\mathrm{tot}}=0.\label{eq:mattcons}
\end{equation}
It is useful to define the density parameter for the mass term (which
will be the effective dark energy density): 
\begin{equation}
\Omega_{\mathrm{mg}}\equiv\frac{\rho_{\mathrm{mg}}}{\rho_{\mathrm{tot}}+\rho_{\mathrm{mg}}}=1-\Omega_{m}-\Omega_{r},
\end{equation}
where $\Omega_{i}=\rho_{i}/(\rho_{\mathrm{tot}}+\rho_{\mathrm{mg}})$
for matter and radiation.

The background dynamics depend entirely on the the $g$-metric Hubble
rate, $\mathcal{H}$, and the ratio of the two scale factors,
$r=b/a$ \cite{2013JHEP...03..099A}. Moreover, by using $N=\log a$
as time variable, with $'$ denoting derivatives with respect to $N$,
the background equations can be conveniently reformulated as a first-order
autonomous system \cite{1475-7516-2014-03-029}: 
\begin{align}
2\mathcal{H}'\mathcal{H}+\mathcal{H}^{2} & =a^{2}(B_{0}+B_{2}r'-w_{\mathrm{tot}}\rho_{\mathrm{tot}}),\\
r' & =\frac{3(1+w_{\mathrm{tot}})B_{1}\Omega_{\mathrm{tot}}r}{\beta_{1}-3\beta_{3}r^{2}-2\beta_{4}r^{3}+3B_{2}r^{2}},\label{eq:rprime}\\
\Omega_{\mathrm{tot}} & =1-\frac{B_{0}}{B_{1}}r,\label{eq:omegam}
\end{align}
where 
\begin{align}
B_{1} & \equiv\beta_{1}+3\beta_{2}r+3\beta_{3}r^{2}+\beta_{4}r^{3},\label{eq:B1}\\
B_{2} & \equiv\beta_{1}+2\beta_{2}r+\beta_{3}r^{2},\label{eq:B2}
\end{align}
and $w_{\mathrm{tot}}$ denotes the equation of state corresponding
to the sum of matter and radiation density parameter $\Omega_{\mathrm{tot}}$.
We can define the effective equation of state 
\begin{align}
w_{\mathrm{eff}} & \equiv\Omega_{\mathrm{mg}}w_{\mathrm{mg}}+\Omega_{\mathrm{tot}}w_{\mathrm{tot}}=-\frac{1}{3}(1+2\frac{\mathcal{H}'}{\mathcal{H}})=-\frac{r(B_{0}+B_{2}r')}{B_{1}}\\
 & =-1+\Omega_{\mathrm{tot}}-\frac{B_{2}rr'}{B_{1}},
\end{align}
from which we obtain
\begin{equation}
w_{\text{mg}}=-1-\frac{B_{2}rr'}{\Omega_{mg}B_{1}}=-1-\frac{B_{2}}{B_{0}}r'.\label{eq:w_mg}
\end{equation}
Another useful relation gives the Hubble rate in terms of $r$ without
an explicit $\rho$ dependence, 
\begin{equation}
\mathcal{H}^{2}=\frac{a^{2}B_{1}}{3r}.
\end{equation}

The background evolution of $r$ will follow Eq. (\ref{eq:rprime})
from an initial value of $r$ until $r'=0$, unless $r$ hits a singularity.
In Ref. \cite{1475-7516-2014-03-029} it was shown that cosmologically
viable evolution can take place in two distinct ways, depending on
initial conditions: when $r$ evolves from 0 to a finite value (we
call this a \textit{finite branch}) and when $r$ evolves from infinity
to a finite value (\textit{infinite branch}). In all viable cases,
the past asymptotic value of $r$ corresponds to $\Omega_{m}=1$ while
the final point corresponds to a de Sitter stage with $\Omega_{m}=0$
(see Fig. \ref{fig:branches} for an illustrative example).

In the following, we consider only pressureless matter, or dust, with
$w_{\mathrm{tot}}=0$. The reason is that we are interested only in
the late-time behavior of bigravity when the Universe is dominated
by dust. We also assume $r\ge0$, although in principle nothing prevents
a negative value of $b$.

We will find it convenient to express all the $\beta_{i}$
parameters in units of $H_{0}^{2}$ and $\mathcal{H}$ in units of
$H_{0}$.%
\footnote{With this convention, our $\beta_{i}$ parameters are equivalent to
the $B_{i}\equiv m^{2}\beta_{i}/H_{0}^{2}$ used in Refs. \cite{2013JHEP...03..099A,2013JCAP...10..046A,2014arXiv1404.4061S}.%
} In this way all the quantities that enter the equations are dimensionless.

\section{Perturbation equations}

\label{sec:perteqs}

In this section we study linear cosmological perturbations. We define
our perturbed metrics in Fourier space by 
\begin{align}
g_{\alpha\beta} & =g_{0,\alpha\beta}+h_{\alpha\beta},\\
f_{\alpha\beta} & =f_{0,\alpha\beta}+h_{f,\alpha\beta},
\end{align}
where $g_{0,\alpha\beta}$ and $f_{0,\alpha\beta}$ are the background
metrics with line elements 
\begin{align}
ds_{g}^{2} & =a^{2}(t)(-dt^{2}+dx_{i}dx^{i}),\\
ds_{f}^{2} & =-[\dot{b}(t)^{2}/\mathcal{H}^{2}(t)]dt^{2}+b(t)^{2}dx_{i}dx^{i},
\end{align}
while $h_{\alpha\beta}$ and $h_{f,\alpha\beta}$ are perturbations
around the backgrounds $g_{0,\alpha\beta}$ and $f_{0,\alpha\beta}$,
respectively, whose line elements are 
\begin{align}
ds_{h}^{2} & =2a^{2}\left[-\Psi dt^{2}+(\Phi\delta_{ij}+k_{i}k_{j}E)dx^{i}dx^{j}\right]\exp(i\mathbf{k}\cdot\mathbf{r}),\\
ds_{h_f}^{2} & =2b^{2}\left[-\frac{\dot{b}^{2}\Psi_{f}}{b^{2}\mathcal{H}^{2}}dt^{2}+(\Phi_{f}\delta_{ij}+k_{i}k_{j}E_{f})dx^{i}dx^{j}\right]\exp(i\mathbf{k}\cdot\mathbf{r}).
\end{align}
After transforming to gauge-invariant variables \cite{Comelli:2012db},
\begin{align}
\Phi & \longrightarrow\Phi-\mathcal{H}^{2}E',\\
\Psi & \longrightarrow\Psi-\mathcal{H}\left(\mathcal{H}'E'+\mathcal{H}\left(E''+E'\right)\right),\\
\Phi_{f} & \longrightarrow\Phi_{f}-\frac{\mathcal{H}{}^{2}rE_{f}'}{r'+r},\\
\Psi_{f} & \longrightarrow\Psi_{f}-\frac{\mathcal{H}r^{2}\mathcal{H}'\left(r'+r\right)E_{f}'+\mathcal{H}^{2}r\left(r\left(r'+r\right)E_{f}''+E_{f}'\left(2r'^{2}+r\left(2r'-r''\right)+r^{2}\right)\right)}{\left(r'+r\right)^{3}},
\end{align}
and using $N=\log a$ as the time variable, the perturbation equations
for the $g_{\mu\nu}$ metric read: 
\begin{align}
[00] & \begin{array}{cc}
\left(\frac{2k^{2}}{3B_{2}a^{2}r}+1\right)\Phi-\Phi_{f}+\frac{1}{3}k^{2}\text{\ensuremath{\Delta E}}+\frac{2H^{3}r\left(-\mathcal{H}+\mathcal{H}'\right)}{\mathcal{A}_{2}}E'-\frac{\mathcal{H}^{2}\mathcal{A}_{1}}{\mathcal{A}_{2}}\text{\ensuremath{\Delta E'}}\end{array}\nonumber \\
 & \begin{array}{cc}
-\frac{2\mathcal{H}^{2}\left(\mathcal{A}_{1}+a^{2}r^{2}B_{2}\right)\left(\mathcal{H}-\mathcal{H}'\right)}{a^{2}k^{2}r\mathcal{A}_{1}B_{2}}\theta-\frac{\delta\rho}{3B_{2}r}=0\end{array},\label{eq:38}\\
{}[0\, i] & \begin{array}{cc}
\Phi'-\Psi+\frac{a^{2}\rho}{2\mathcal{H}k^{2}}\theta+\left(\mathcal{H}^{2}-\mathcal{H}\mathcal{H}'\right)E'=0,\end{array}\label{eq:0i-g}\\
{}[i\: j] & \begin{array}{cc}
\Phi+\Psi+\frac{1}{2}a^{2}r\mathcal{A}_{3}\text{\ensuremath{\Delta E=0,}}\end{array}\\
{}[i\,\, i] & \begin{array}{cc}
\left(\frac{2k^{2}}{3B_{2}a^{2}r}+\frac{\mathcal{A}_{3}}{B_{2}}\right)\Phi+\left(\frac{2k^{2}}{3B_{2}a^{2}r}+1\right)\Psi-\frac{\mathcal{A}_{3}}{B_{2}}\Phi_{f}-\frac{\mathcal{A}_{2}}{\mathcal{A}_{1}}\Psi_{f}+\frac{k^{2}\mathcal{A}_{3}}{3B_{2}}\Delta E-\frac{2\mathcal{H}^{3}r\left(\mathcal{H}-\mathcal{H}'\right)}{\mathcal{A}_{2}}E''\end{array}\nonumber \\
 & \begin{array}{cc}
-\frac{\mathcal{H}^{2}\mathcal{A}_{1}}{\mathcal{A}_{2}}\text{\ensuremath{\Delta E''}}+\mathcal{A}_{4}E'+\mathcal{A}_{5}\Delta E'=0,\end{array}
\end{align}
while the corresponding equations for $f_{\mu\nu}$ are 
\begin{align}
[00] & \begin{array}{cc}
\Phi-\left(1+\frac{2k^{2}r}{3a^{2}B_{2}}\right)\Phi_{f}+\frac{k^{2}}{3}\text{\text{\ensuremath{\Delta E}}}-\frac{\mathcal{A}_{1}\mathcal{H}^{2}}{\mathcal{A}_{2}}\text{\ensuremath{\Delta E}}'-\frac{2\mathcal{H}^{3}r\left(\mathcal{H}-\mathcal{H}'\right)}{\mathcal{A}_{2}}E'=0,\end{array}\\
{}[0\, i] & \begin{array}{cc}
\Phi_{f}'-\frac{\mathcal{A}_{2}}{\mathcal{A}_{1}}\Psi_{f}+\frac{a^{2}\mathcal{H}B_{2}\left(\mathcal{H}'-\mathcal{H}\right)}{\mathcal{A}_{2}}\text{\ensuremath{\Delta E'}}-\frac{a^{2}\mathcal{H}B_{2}\left(\mathcal{H}'-\mathcal{H}\right)}{\mathcal{A}_{2}}E'=0,\end{array}\label{eq:0i-f}\\
{}[i\: j] & \begin{array}{cc}
\Phi_{f}+\Psi_{f}-\frac{a^{2}\mathcal{A}_{1}\mathcal{A}_{3}}{2r\mathcal{A}_{2}}\text{\ensuremath{\Delta E}}=0,\end{array}\\
{}[i\,\, i] & \begin{array}{cc}
\left(\frac{2rk^{2}\mathcal{A}_{2}}{3a^{2}B_{2}\mathcal{A}_{1}}+\frac{\mathcal{A}_{3}}{B_{2}}\right)\Phi_{f}+\left(\frac{2k^{2}r\mathcal{A}_{2}}{3a^{2}B_{2}\mathcal{A}_{1}}+\frac{\mathcal{A}_{2}}{\mathcal{A}_{1}}\right)\Psi_{f}-\frac{\mathcal{A}_{3}}{B_{2}}\Phi-\Psi-\frac{k^{2}\mathcal{A}_{3}}{3B_{2}}\text{\ensuremath{\Delta E}}+\frac{2\mathcal{H}^{3}r\left(\mathcal{H}'-\mathcal{H}\right)}{\mathcal{A}_{2}}E''\end{array}\nonumber \\
{} & \begin{array}{cc}
+\frac{\mathcal{H}^{2}\mathcal{A}_{1}}{\mathcal{A}_{2}}\text{\ensuremath{\Delta E''}}-\mathcal{A}_{4}E'-\mathcal{A}_{5}\Delta E'=0,\end{array}\label{eq:45}
\end{align}
where $\Delta E\equiv E-E_{f}$ and the $\mathcal{A}_{i}$ coefficients
are defined as 
\begin{align}
\mathcal{A}_{1} & =a^{2}B_{2}-2\mathcal{H}^{2}r, \\
\mathcal{A}_{2} & =a^{2}B_{2}-2\mathcal{H}r\mathcal{H}', \\
\mathcal{\mathcal{A}}_{3} & =2B_{2}+B_{2}', \\
\mathcal{\mathcal{A}}_{4} & =-\frac{\left(\mathcal{A}_{1}-\mathcal{A}_{2}\right){}^{2}\left(-a^{4}\left(1+2r^{2}\right)B_{2}^{2}+\mathcal{A}_{1}\left(\mathcal{A}_{1}+\mathcal{A}_{2}\right)+a^{2}r^{2}B_{2}\left(2\mathcal{A}_{1}+\mathcal{A}_{2}\right)\right)}{2r\left(a^{2}r^{2}B_{2}+\mathcal{A}_{1}\right)\mathcal{A}_{2}^{2}}\nonumber \\
 & \hphantom{{}=}+\frac{\left(-a^{2}B_{2}+\mathcal{A}_{1}\right)\left(\mathcal{A}_{1}-\mathcal{A}_{2}\right)\left(\mathcal{A}_{1}\mathcal{A}_{2}-a^{2}B_{2}\left(\left(1+r^{2}\right)\mathcal{A}_{1}-r^{2}\mathcal{A}_{2}\right)\right)B_{2}'}{2rB_{2}\left(a^{2}r^{2}B_{2}+\mathcal{A}_{1}\right)\mathcal{A}_{2}^{2}}, \\
\mathcal{\mathcal{A}}_{5} & =\frac{\mathcal{A}_{1}^{2}\left(\mathcal{A}_{1}^{2}-\mathcal{A}_{1}\mathcal{A}_{2}-4\mathcal{A}_{2}^{2}\right)+a^{2}B_{2}\mathcal{A}_{1}\left(2r^{2}\mathcal{A}_{1}^{2}-3r^{2}\mathcal{A}_{1}\mathcal{A}_{2}+\left(4-3r^{2}\right)\mathcal{A}_{2}^{2}\right)}{2r\left(a^{2}r^{2}B_{2}+\mathcal{A}_{1}\right)\mathcal{A}_{2}^{2}}\nonumber \\
 &  \hphantom{{}=}-\frac{a^{4}B_{2}^{2}\left(\left(1+2r^{2}\right)\mathcal{A}_{1}^{2}-2\left(1+2r^{2}\right)\mathcal{A}_{1}\mathcal{A}_{2}+\left(1-2r^{2}\right)\mathcal{A}_{2}^{2}\right)}{2r\left(a^{2}r^{2}B_{2}+\mathcal{A}_{1}\right)\mathcal{A}_{2}^{2}}\nonumber \\
 &  \hphantom{{}=}+\frac{\mathcal{A}_{1}\left(-a^{2}B_{2}+\mathcal{A}_{1}\right)\left(-\mathcal{A}_{1}\mathcal{A}_{2}+a^{2}B_{2}\left(\left(1+r^{2}\right)\mathcal{A}_{1}-\left(1+2r^{2}\right)\mathcal{A}_{2}\right)\right)B_{2}'}{2rB_{2}\left(a^{2}r^{2}B_{2}+\mathcal{A}_{1}\right)\mathcal{A}_{2}^{2}}.
\end{align}
These equations are in agreement with those presented in Refs. \cite{2012JCAP...12..021B,Comelli:2012db,2014arXiv1404.4061S} (for a more detailed derivation see, e.g., Ref. \cite{2012PhRvD..86d3517K}).

The matter equations are 
\begin{equation}
\delta'+\theta\mathcal{H}^{-1}+3\Phi'-3\mathcal{H}^{2}E''-6\mathcal{H}\mathcal{H}'E'+k^{2}E'=0,\label{eq:perteq7_mm}
\end{equation}
\begin{equation}
\theta'+\theta+k^{2}E'\mathcal{H}'-k^{2}\Psi\mathcal{H}^{-1}+k^{2}\mathcal{H}\left(E''+E'\right)=0,\label{eq:perteq8_mm}
\end{equation}
where $\delta$ and $\theta$ are the matter density contrast and peculiar
velocity divergence, respectively. Differentiating and combining Eqs. (\ref{eq:perteq7_mm})
and (\ref{eq:perteq8_mm}) we obtain 
\begin{equation}
\delta''+\left(1+\frac{\mathcal{H}'}{\mathcal{H}}\right)\delta'+\frac{k^{2}\Psi}{\mathcal{H}^{2}}-6E'\left(2\mathcal{H}'^{2}+\mathcal{H}\left(\mathcal{H}''+\mathcal{H}'\right)\right)-3\mathcal{H}E''\left(5\mathcal{H}'+\mathcal{H}\right)-3E^{(3)}\mathcal{H}^{2}+3\left(1+\frac{\mathcal{H}'}{\mathcal{H}}\right)\Phi'+3\Phi''=0.\label{eq:delta_full}
\end{equation}
Note that $E$ enters the equations only with derivatives; one could
then define a new variable $Z=E'$ to lower the degree of the equations.%
\footnote{$E$ only appears without derivatives in the mass terms, specifically in differences
with $E_{f}$, and so all appearances of $E$ are accounted for by the separate gauge-invariant
variable $\Delta E$.%
} One could also adopt the gauge-invariant variables 
\begin{align}
\delta & \to\delta+3\mathcal{H}^{2}E',\\
\theta & \to\theta-k^{2}\mathcal{H}E'
\end{align}
to bring the matter conservation equations into the standard form
of a longitudinal gauge but since this renders the other equations
somewhat more complicated we will not employ them.

\section{Quasistatic limit}

Large-scale structure experiments predominantly probe modes within
the horizon. Conveniently, in the subhorizon and quasistatic limit,
the cosmological perturbation equations simplify dramatically. In
this section we consider this QSlimit of subhorizon
structures in bimetric gravity.

The subhorizon limit is defined by assuming $k\gg\mathcal{H}$, while
the QS limit assumes that modes oscillate on a Hubble timescale:
$\Xi'\sim\Xi$ for any variable $\Xi$.%
\footnote{Recall that we are using the dimensionless $N=\log a$ as our time
variable.} Concretely, this means that we consider the regime where $(k^{2}/\mathcal{H}^{2})\Xi_{i}\gg\Xi_{i}\sim\Xi_{i}'\sim\Xi_{i}''$
for each field $\Xi_{i}=\{\Psi,\Phi,\Psi_{f},\Phi_{f},\Delta E,E\}$.
We additionally take $\delta(k/\mathcal{H})^{2},\delta'(k/\mathcal{H})^{2}\gg\theta/\mathcal{H}$.
In this limit we obtain the system of equations 
\begin{align}
 & \begin{array}{cc}
3k^{2}\text{\ensuremath{\Delta E}+}\left(9+\frac{6k^{2}}{B_{2}a^{2}r}\right)\Phi-9\Phi_{f}-\frac{3\delta\rho}{B_{2}r}=0,\end{array}\\
{} & \begin{array}{cc}
\frac{1}{2}a^{2}r\mathcal{A}_{3}\Delta E+\Phi+\Psi\text{\ensuremath{=0,}}\end{array}\\
{} & 3\frac{k^{2}\mathcal{A}_{3}}{B_{2}}\Delta E+\begin{array}{cc}
\left(9\frac{\mathcal{A}_{3}}{B_{2}}+\frac{6k^{2}}{B_{2}a^{2}r}\right)\Phi+\left(9+\frac{6k^{2}}{B_{2}a^{2}r}\right)\Psi-9\frac{\mathcal{A}_{3}}{B_{2}}\Phi_{f}-9\frac{\mathcal{A}_{2}}{\mathcal{A}_{1}}\Psi_{f}=0,\end{array}\\
{} & \begin{array}{cc}
3k^{2}\text{\text{\ensuremath{\Delta E}}}-\left(9+\frac{6k^{2}r}{a^{2}B_{2}}\right)\Phi_{f}+9\Phi=0,\end{array}\\
{} & \begin{array}{cc}
-\frac{a^{2}\mathcal{A}_{1}\mathcal{A}_{3}}{2r\mathcal{A}_{2}}\text{\ensuremath{\Delta E}}+\Phi_{f}+\Psi_{f}=0,\end{array}\\
{} & \frac{3k^{2}\mathcal{A}_{3}}{B_{2}}\text{\ensuremath{\Delta E}}+\frac{9\mathcal{A}_{3}}{B_{2}}\Phi+9\Psi-\left(\frac{6rk^{2}\mathcal{A}_{2}}{a^{2}B_{2}\mathcal{A}_{1}}+\frac{9\mathcal{A}_{3}}{B_{2}}\right)\Phi_{f}-\left(\frac{6k^{2}r\mathcal{A}_{2}}{a^{2}B_{2}\mathcal{A}_{1}}+\frac{9\mathcal{A}_{2}}{\mathcal{A}_{1}}\right)\Psi_{f}=0,
\end{align}
where we have used the momentum constraints, Eqs. (\ref{eq:0i-g})
and (\ref{eq:0i-f}), to replace time derivatives of $\Phi$ and $\Phi_{f}$.
The above set of equations can be solved for $\Psi,\Phi,\Psi_{f},\Phi_{f}$,
and $\Delta E$ in terms of $\delta$ (see also Ref. \cite{2014arXiv1404.4061S}):
\begin{align}
\Psi & =\frac{3\left(3a^{2}\mathcal{A}_{1}\mathcal{A}_{3}B_{2}^{2}+3a^{2}\mathcal{A}_{2}\mathcal{A}_{3}B_{2}^{2}r^{2}+k^{2}\left(2\mathcal{A}_{1}\mathcal{A}_{3}^{2}r^{3}-2B_{2}r\left(\mathcal{A}_{2}B_{2}-2\mathcal{A}_{1}\mathcal{A}_{3}\right)\right)\right)\Omega_{m}\mathcal{H}^{2}}{k^{4}\left(B_{2}^{2}\left(4\mathcal{A}_{1}r^{3}+4\mathcal{A}_{2}r\right)-8\mathcal{A}_{1}\mathcal{A}_{3}B_{2}r\left(r^{2}+1\right)\right)-6k^{2}\left(r^{2}+1\right)^{2}a^{2}\mathcal{A}_{1}\mathcal{A}_{3}B_{2}^{2}}\delta,\label{eq:solQS_Psi-gen}\\
\Phi & =-\frac{3\left(3a^{2}\mathcal{A}_{1}\mathcal{A}_{3}B_{2}+3a^{2}\mathcal{A}_{1}\mathcal{A}_{3}B_{2}r^{2}+k^{2}\left(r\left(4\mathcal{A}_{1}\mathcal{A}_{3}-2\mathcal{A}_{2}B_{2}\right)+2\mathcal{A}_{1}\mathcal{A}_{3}r^{3}\right)\right)\Omega_{m}\mathcal{H}^{2}}{k^{4}\left(B_{2}\left(4\mathcal{A}_{1}r^{3}+4\mathcal{A}_{2}r\right)-8\mathcal{A}_{1}\mathcal{A}_{3}r\left(r^{2}+1\right)\right)-6k^{2}\left(r^{2}+1\right)^{2}a^{2}\mathcal{A}_{1}\mathcal{A}_{3}B_{2}}\delta,\label{eq:solQS_Phi-gen}\\
\Psi_{f} & =-\frac{3\left(-3a^{4}\mathcal{A}_{1}^{2}\mathcal{A}_{3}B_{2}^{2}-3a^{4}\mathcal{A}_{1}\mathcal{A}_{2}\mathcal{A}_{3}B_{2}^{2}r^{2}+2\mathcal{A}_{1}k^{2}r\left(a^{2}\mathcal{A}_{1}\mathcal{A}_{3}^{2}-a^{2}\left(\mathcal{A}_{1}+\mathcal{A}_{2}\right)\mathcal{A}_{3}B_{2}+\mathcal{A}_{2}B_{2}^{2}\right)\right)\Omega_{m}\mathcal{H}^{2}}{k^{4}\left(B_{2}^{2}\left(4\mathcal{A}_{1}\mathcal{A}_{2}r^{3}+4\mathcal{A}_{2}^{2}r\right)-8\mathcal{A}_{1}\mathcal{A}_{2}\mathcal{A}_{3}B_{2}r\left(r^{2}+1\right)\right)-6k^{2}\left(r^{2}+1\right)^{2}a^{2}\mathcal{A}_{1}\mathcal{A}_{2}\mathcal{A}_{3}B_{2}^{2}}\delta,\label{eq:solQS_Psif-gen}\\
\Phi_{f} & =-\frac{3\left(3a^{2}\mathcal{A}_{1}\mathcal{A}_{3}B_{2}+3a^{2}\mathcal{A}_{1}\mathcal{A}_{3}B_{2}r^{2}+2\mathcal{A}_{1}k^{2}r\left(\mathcal{A}_{3}-B_{2}\right)\right)\Omega_{m}\mathcal{H}^{2}}{k^{4}\left(B_{2}\left(4\mathcal{A}_{1}r^{3}+4\mathcal{A}_{2}r\right)-8\mathcal{A}_{1}\mathcal{A}_{3}r\left(r^{2}+1\right)\right)-6k^{2}\left(r^{2}+1\right)^{2}a^{2}\mathcal{A}_{1}\mathcal{A}_{3}B_{2}}\delta,\label{eq:solQS_Phif-gen}\\
\Delta E & =\frac{3r\left(3a^{2}\left(\mathcal{A}_{1}-\mathcal{A}_{2}\right)B_{2}^{2}+2\mathcal{A}_{1}k^{2}r\left(B_{2}-\mathcal{A}_{3}\right)\right)\Omega_{m}\mathcal{H}^{2}}{k^{4}a^{2}\left(B_{2}^{2}\left(2\mathcal{A}_{1}r^{3}+2\mathcal{A}_{2}r\right)-4\mathcal{A}_{1}\mathcal{A}_{3}B_{2}r\left(r^{2}+1\right)\right)-3a^{2}k^{2}\left(r^{2}+1\right)^{2}a^{2}\mathcal{A}_{1}\mathcal{A}_{3}B_{2}^{2}}\delta.\label{eq:solQS_DeltaE-gen}
\end{align}

The QS limit is, however, only a good approximation if the full set
of equations produces a stable solution for large $k$. In fact, if
the solutions are not stable, the derivative terms we have neglected
are no longer small (as their mean values vary on a faster timescale
than Hubble), and the QS limit is never reached. We therefore need
to analyze the stability of the full theory.

\section{Instabilities}

Let us go back to the full linear equations, presented in section~\ref{sec:perteqs}.
While we have ten equations for ten variables, there are only two
independent degrees of freedom, corresponding to the scalar modes
of the two gravitons. The degrees-of-freedom counting goes as follows
(see Ref. \cite{2014PhRvD..89b4034L} for an in-depth discussion of
most of these points): four of the metric perturbations ($\delta g_{00}$,
$\delta g_{0i}$, $\delta f_{00}$, and $\delta f_{0i}$) and $\theta$
are nondynamical, as their derivatives do not appear in the second-order
action. These can be integrated out in terms of the dynamical variables
and their derivatives. We can further gauge fix two of the dynamical
variables. Finally, after the auxiliary variables are integrated out, one of the initially dynamical variables \textit{becomes} auxiliary (its derivatives drop out of the action) and can itself be integrated out.\footnote{We thank Macarena Lagos and Pedro Ferreira for discussions on this point.}

This leaves us with two independent dynamical degrees of freedom.
The aim of this section is to reduce the ten linearized Einstein equations
to two coupled second-order equations, and then ask whether the solutions
to that system are stable. We will choose to work with $\Phi$ and
$\Psi$ as our independent variables, eliminating all of the other
perturbations in their favor.

We can begin by eliminating $\Psi_{f}$, $\Phi_{f}$, $\Delta E$,
and their derivatives using the $0-0$, $i-i$, and $i-j$ components
of the $g$-metric perturbation equations. We will herein refer to
these equations as $g_{00}$, $g_{ii}$, and so on for the sake of
conciseness. Doing this we see also that the $g_{ij}$ and $f_{ij}$
equations are linearly related. Then we can replace $\delta$ and
$\theta$ with the help of the $g_{0i}$ and $f_{00}$ equations.
Finally, one can find a linear combination of the $f_{0i}$ and $g_{ii}$
equations which allows one to express $E'$ as a function of $\Phi$,
$\Psi$, and their derivatives. In this way, we can write our original
ten equations as just two second-order equations for $X_{i}\equiv\{\Phi,\Psi\}$
with the following structure: 
\begin{equation}
X_{i}''+F_{ij}X'_{j}+S_{ij}X_{j}=0,\label{eq:2nd_order_DEQ}
\end{equation}
where $F_{ij}$ and $S_{ij}$ are complicated expressions that depend
only on background quantities and on $k$. The eigenfrequencies of
these equations can easily be found by substituting $X=X_{0}e^{i\omega N}$,
assuming that the dependence of $\omega$ on time is negligibly small.%
\footnote{The criterion for this WKB approximation to hold is $|\omega'/\omega^{2}|\ll1$.
We find that for large $k$ this approximation is almost always valid.%
} For instance, assuming that only $\beta_{1}$ is nonzero, in the
limit of large $k$ we find \cite{Konnig:2014dna} 
\begin{equation}
\omega_{\beta_{1}}=\pm\frac{k}{\mathcal{H}}\frac{\sqrt{-1+12r^{2}+9r^{4}}}{1+3r^{2}},\label{eq:eigenfreq_b1}
\end{equation}
plus two other solutions that are independent of $k$ and are therefore
subdominant. One can see then that real solutions (needed to obtain oscillating, rather than growing and decaying, solutions for $X$) are found
only for $r>0.28$, which occurs for $N=-0.4$, i.e., $z\approx0.5$.
At any epoch before this, the perturbation equations are unstable
for large $k$. In other words, we find an imaginary sound speed.
This behavior invalidates linear perturbation theory
on subhorizon scales and may rule out the model, if the instability
is not cured at higher orders, for instance by a phenomenology related
to the Vainshtein mechanism \cite{Vainshtein:1972sx,Babichev:2013usa}.

Now let us move on to more general models. Although the other one-parameter
models are not viable in the background\footnote{With the exception of the $\beta_{0}$ model, which is simply $\Lambda$CDM.%
} (i.e., none of them have a matter dominated epoch in the asymptotic
past and produce a positive Hubble rate) \cite{1475-7516-2014-03-029},
it is worthwhile to study the eigenfreqencies in these cases too,
particularly because they will tell us the early time behavior of
the viable multiple-parameter models. For simplicity, from now on we refer
to a model in which, e.g., only $\beta_{1}$ and $\beta_{2}$ are
nonzero as the $\beta_{1}\beta_{2}$ model, and so on.

At early times, every viable, finite-branch, multiple-parameter model
reduces to the single-parameter model with the lowest-order interaction.
For instance, the $\beta_{1}\beta_{2}$, $\beta_{1}\beta_{3}$, and
$\beta_{1}\beta_{2}\beta_{3}$ models all reduce to $\beta_{1}$,
the $\beta_{2}\beta_{3}$ model reduces to $\beta_{2}$, and so on.
Similarly, in the early Universe, the viable, infinite-branch models
reduce to single-parameter models with the highest-order interaction.
Therefore, in order to determine the early time stability, we need
to only look at the eigenfrequencies of single-parameter models, for
which we find
\begin{align}
\omega_{\beta_{2}} & =\pm\frac{k}{\mathcal{H}r},\label{eq:eigenfreq_b2}\\
\omega_{\beta_{3}} & =\pm\frac{ik\sqrt{r^{4}-8r^{2}+3}}{\sqrt{3}\mathcal{H}\left(r^{2}-1\right)},\label{eq:eigenfreq_b3}\\
\omega_{\beta_{4}} & =\pm\frac{k}{\sqrt{2}\mathcal{H}}.\label{eq:eigenfreq_b4}
\end{align}
Therefore, the only single-parameter models without instabilities
at early times are the $\beta_{2}$ and $\beta_{4}$ models. Using
the rules discussed above, we can now extend these results to the rest
of the bigravity parameter space.

Since much of the power of bigravity lies in its potential
to address the dark energy problem in a technically natural way, let
us first consider models without an explicit $g$-metric cosmological
constant, i.e., $\beta_{0}=0$. On the finite branch, all such models
with $\beta_1 \neq0$ reduce, at early times, to the $\beta_{1}$ model,
which has an imaginary eigenfrequency for large $k$ (\ref{eq:eigenfreq_b1})
and is therefore unstable in the early Universe. Hence the finite-branch
$\beta_{1}\beta_{2}\beta_{3}\beta_{4}$ model and its subsets
with $\beta_{1}\neq0$ are all plagued by instabilities. All of these models have viable
background evolution \cite{1475-7516-2014-03-029}. This leaves the $\beta_{2}\beta_{3}\beta_{4}$
model; this is stable on the finite branch as long as $\beta_{2}\neq0$,
but its background is not viable. We conclude that there
are no models with $\beta_{0}=0$ which live on a finite branch, have a viable background evolution,
and predict stable linear perturbations at all times.

This conclusion has two obvious loopholes: either including a cosmological
constant, $\beta_{0}$, or turning to an infinite-branch model. We first consider including a nonzero cosmological constant, although this may not be as interesting theoretically as the
models which self accelerate. Adding a cosmological constant can change the stability properties, although it turns out not to do so in the finite-branch models with viable backgrounds. In the $\beta_{0}\beta_{1}$ model, the eigenfrequencies, 
\begin{equation}
\omega_{\beta_{0}\beta_{1}}=\pm\frac{k\sqrt{9r^{4}+2\left(\beta_{0}/\beta_{1}\right)r+12r^{2}-1}}{\mathcal{H}\left(3r^{2}+1\right)},
\end{equation}
are unaffected by $\beta_{0}$ at early times and therefore still imply unstable
modes in the asymptotic past. This result extends (at early times) to the rest of the bigravity parameter space with $\beta_0,\beta_1\neq0$. No other finite-branch models yield viable backgrounds. Therefore, all of the
solutions on a finite branch, for any combination of parameters, are
either unviable (in the background) or linearly unstable in the past.

Let us now turn to the infinite-branch models. In this case, it turns
out that there exists a small class of viable models which has stable
cosmological evolution: models where the only nonvanishing parameters
are $\beta_{0}$, $\beta_{1}$, and $\beta_{4}$, as well as the self-accelerating
$\beta_{1}\beta_{4}$ model. Here, $r$ evolves from infinity in the
past and asymptotes to a finite de Sitter value in the future. As mentioned in Ref.~\cite{1475-7516-2014-03-029}, a nonvanishing $\beta_2$ or $\beta_3$ would not be compatible with the requirement $\lim_{t\rightarrow-\infty}\Omega_{tot}=1$. This can be seen directly from Eq. (\ref{eq:omegam}) in the limit of large $r$. For
these $\beta_{0}\beta_{1}\beta_{4}$ models we perform a similar eigenfrequency
analysis and obtain 
\begin{align}
\omega_{\beta_{0}\beta_{1}\beta_{4}} & =\pm\frac{k\sqrt{\left(9+2\beta_{0}\beta_{4}/\beta_{1}^{2}\right)r^{4}+2\left(\beta_{0}/\beta_{1}\right)r+12r^{2}-1+\left(\beta_{4}/\beta_{1}\right)\left[2(\beta_{4}/\beta_{1})r^{6}-6r^{5}-8r^{3}\right]}}{\mathcal{H}\left(3r^{2}+1-2\left(\beta_{4}/\beta_{1}\right)r^{3}\right)}.\label{eq:eqigenfreq_b0b1b4}
\end{align}
Restricting ourselves to the self-accelerating models (i.e., $\beta_{\text{0}}=0$), we obtain 
\begin{align}
\omega_{\beta_{1}\beta_{4}} & =\pm\frac{k\sqrt{9r^{4}+12r^{2}-1+\left(\beta_{4}/\beta_{1}\right)\left[2(\beta_{4}/\beta_{1})r^{6}-6r^{5}-8r^{3}\right]}}{\mathcal{H}\left(3r^{2}+1-2\left(\beta_{4}/\beta_{1}\right)r^{3}\right)}.\label{eq:eigenfreq_b1b4}
\end{align}
Notice that, for large $r$, the eigenvalues (\ref{eq:eqigenfreq_b0b1b4})-(\ref{eq:eigenfreq_b1b4}) reduce to the expression (\ref{eq:eigenfreq_b4})
for $\omega_{\beta_{4}}$. This frequency is real, and therefore the
$\beta_{1}\beta_{4}$ model, as well as its generalization to include
a cosmological constant, is stable on the infinite branch at early times.

\begin{figure}
\includegraphics[width=0.6\textwidth]{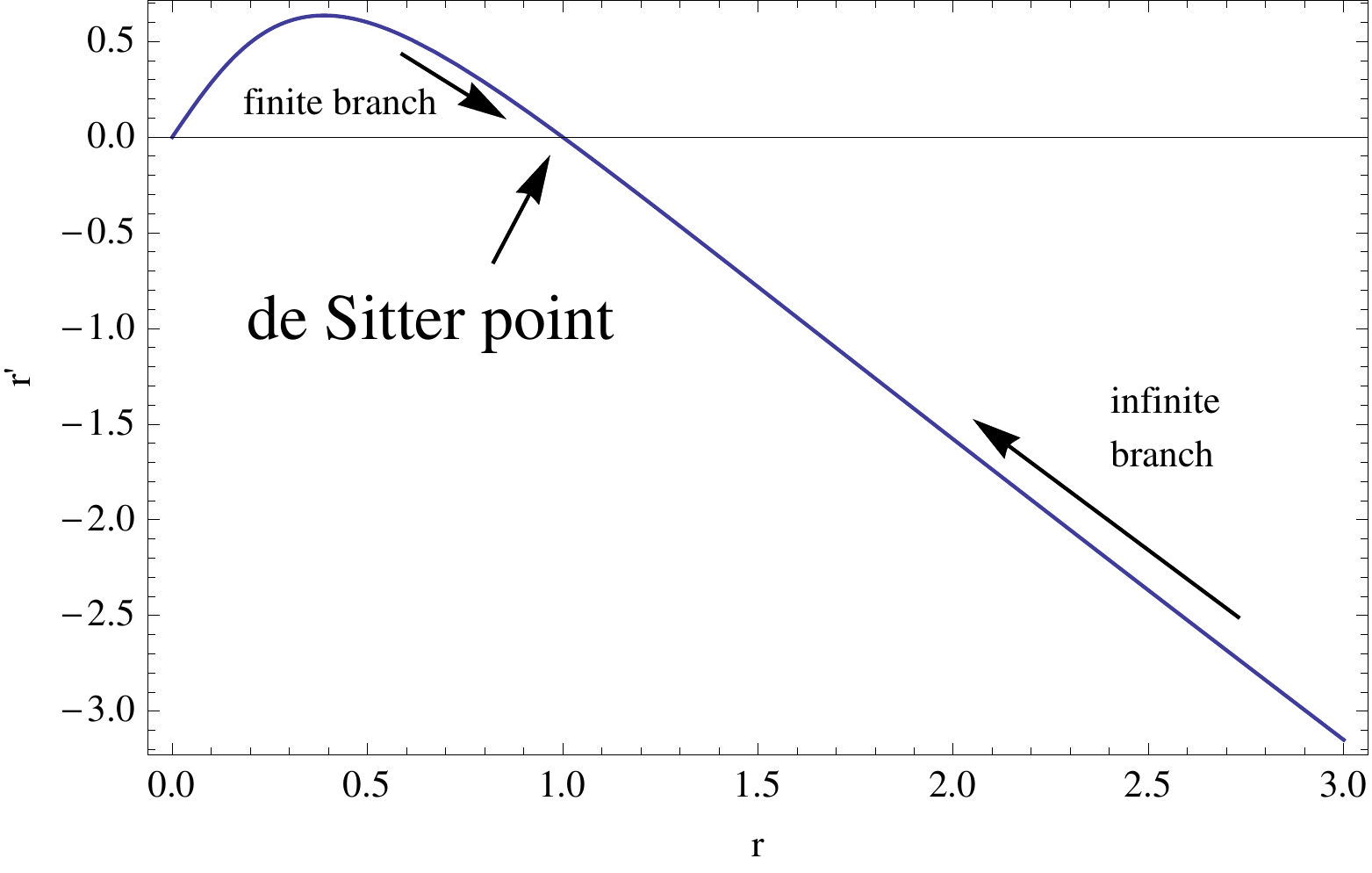}

\protect\protect\caption{\label{fig:branches}Plot of the function $r'(r)$ for the $\beta_{1}\beta_{4}$
model for $\beta_{1}=0.5$, $\beta_{4}=1$. For both the finite and infinite
branches, the final state is the de Sitter point. The arrows show
the direction of movement of $r$.}
\end{figure}

It is interesting to note that the eigenfrequencies can also be written
as 
\begin{align}
\omega_{\beta_{0}\beta_{1}\beta_{4}} & =\pm\frac{ik}{\mathcal{H}}\sqrt{\frac{r''}{3r'}}.\label{eq:eqigenfreq_b0b1b4-1}
\end{align}
Therefore, the condition for the stability of this model in the infinite
branch, where $r'<0$, is simply $r''>0$. One might wonder whether
this expression for $\omega$ is general or model specific. While it does not hold for the $\beta_{2}$ and $\beta_{3}$ models, Eqs.
(\ref{eq:eigenfreq_b2}) and (\ref{eq:eigenfreq_b3}), it is valid for all of the submodels of $\beta_{0}\beta_{1}\beta_{4}$,
including Eqs. (\ref{eq:eigenfreq_b1})-(\ref{eq:eigenfreq_b4}). We can see
from this, for example, that the finite-branch ($r'>0$) $\beta_{1}$
model is unstable at early times because initially $r''$ is positive.
In Fig. \ref{fig:branches} we show schematically the evolution of
the $\beta_{1}\beta_{4}$ model on the finite and infinite branches.
The stability condition on either branch is $r''/r'=dr'/dr<0$. For the parameters plotted, $\beta_1=0.5$ and $\beta_4=1$, one can see graphically that this condition is met, and hence the model is stable, only at late times on the finite branch but for all times on the infinite branch. Our remaining task is to extend this to other parameters.

Let us now prove that the infinite-branch $\beta_{1}\beta_{4}$ model is
stable at all times for all viable choices of the parameters. In a previous work we showed that background viability and the condition
that we live on the infinite branch restrict us to the parameter range
$0<\beta_{4}<2\beta_{1}$ \cite{1475-7516-2014-03-029,2014arXiv1404.4061S}. We have already seen that at early times, $r\rightarrow\infty$,
and the eigenfrequencies match those in the $\beta_{4}$ model (\ref{eq:eigenfreq_b4})
which are purely real. What about later times? The discriminant is
positive and hence the model is stable whenever $r>1$. The question then is the following: do the infinite-branch models in this region
of the parameter space always have $r>1$?

The answer is \textit{yes}. To see this, consider the algebraic equation
for $r$, which can be determined by combining the $g$- and $f$-metric
Friedmann equations (see Eq. (2.17) of Ref.~\cite{2014arXiv1404.4061S}),
and focus on the asymptotic future by taking $\rho\to0$. This gives
\begin{equation}
\beta_{4}r_{c}^{3}-3\beta_{1}r_{c}^{2}+\beta_{1}=0,\label{eq:rcubic}
\end{equation}
where $r_{c}$ is the far-future value of $r$. When $\beta_{4}=2\beta_{1}$
exactly, this is solved by $r_{c}=1$. We must then ask whether for
$0<\beta_{4}<2\beta_{1}$, $r_{c}$ remains greater than 1. Writing
$p\equiv r_{c}-1$, using Descartes' rule of signs, and restricting
ourselves to $0<\beta_{4}<2\beta_{1}$, we can see that $p$ has
one positive root, i.e., there is always exactly one solution with $r_{c}>1$ in that
parameter range. Therefore, in all infinite-branch solutions with
$0<\beta_{4}<2\beta_{1}$, $r$ evolves to some $r_{c}>1$ in the
asymptotic future. We conclude that all of the infinite-branch $\beta_{1}\beta_{4}$
cosmologies which are viable at the background level are also linearly
stable at all times, providing a clear example of a bimetric cosmology which is a viable competitor to $\Lambda$CDM.
\par
The models without quadratic- and cubic-order interactions were also discussed in Ref.~\cite{Hassan:2014vja}. Interestingly, for those models, as well as other models where only one of the three parameters $\beta_1$, $\beta_2$ and $\beta_3$ is nonvanishing, the authors found that if one metric is an Einstein metric, i.e. $G_{\mu\nu} + \Lambda g_{\mu\nu} = 0$, then the other metric is proportional to it. This automatically avoids pathologic solutions when choosing the nondynamical constraint in the Bianchi constraint \cite{Hassan:2014vja} (which are, however, explicitly avoided in the present work by imposing the dynamical constraint in order to find cosmological solutions that differ from $\Lambda$CDM).

\section{Quasistatic limit of infinite-branch bigravity}

In the previous section we found that most bigravity models which
are viable at the background level suffer from a linear instability
at early times. A prominent exception was the model with the $\beta_{1}$
and $\beta_{4}$ interactions turned on (i.e., the first-order interaction between the two metrics
and the $f$-metric cosmological constant) in the case of
solutions on the \textit{infinite branch}, where $r$ evolves
from infinity at early times to a finite value in the far future.
This means that we can safely use the QS approximation for the subhorizon
modes in the infinite-branch $\beta_{1}\beta_{4}$ model, hereafter referred to (interchangeably) as infinite-branch bigravity (IBB); in this
section, we compare the QS limit of this model to observations.

The background cosmology of IBB was studied in Refs.~\cite{1475-7516-2014-03-029,2014arXiv1404.4061S}.
Reference~\cite{2014arXiv1404.4061S} further studied the linear perturbations
and quasistatic limit, finding results in agreement with those presented
in the following two sections. Using the Friedmann equations, it has
been shown that the background cosmology only selects a curve in the
parameter space, given by
\begin{equation}
\beta_{4}=\frac{3\Omega_{\mathrm{mg,0}}\beta_{1}^{2}-\beta_{1}^{4}}{\Omega_{\mathrm{mg,0}}^{3}},\label{eq:degeneracy_curve_b1b4}
\end{equation}
where we recall that $\Omega_{\mathrm{mg,0}}\equiv\beta_{1}r_{0}$
is the present-day effective density of dark energy that appears
in the Friedmann equation (\ref{eq:fried-1}). This does not need
to coincide with the value of $\Omega_{\Lambda}$ derived in the context
of $\Lambda$CDM models; indeed, the best-fit value to the background
data is $\Omega_{\mathrm{mg,0}}=0.84_{-0.02}^{+0.03}$ \cite{1475-7516-2014-03-029}.
Furthermore, as discussed in the previous subsection, to ensure that
we are on the infinite branch we impose the condition $0<\beta_{4}<2\beta_{1}$.

The QS-limit equations in terms of $\delta$ now read (recall $B_{1}=\beta_{1}+\beta_{4}r^{3}$,
see Eq. (\ref{eq:B1})): 
\begin{align}
k^{2}\Psi & =\frac{\left(\frac{3}{2}a^{2}\beta_{1}\left(9\beta_{1}\left(r^{2}-1\right)r^{2}+\left(r^{2}-2\right)\mathcal{B}\right)-\frac{1}{2}k^{2}r\left(9\beta_{1}\left(r^{2}-1\right)+\left(8r^{2}+9\right)\mathcal{B}\right)\right)\Omega_{m}\mathcal{H}^{2}}{3a^{2}\beta_{1}\left(r^{2}+1\right)^{2}\mathcal{B}+k^{2}\left(2r^{3}\mathcal{B}+3\beta_{1}\left(r^{2}-1\right)r+3r\mathcal{B}\right)}\delta,\label{eq:solQS_Psi-gen-1}\\
k^{2}\Phi & =\frac{\left(3a^{2}\beta_{1}\left(r^{2}+1\right)\mathcal{B}+\frac{1}{2}k^{2}r\left(9\beta_{1}\left(r^{2}-1\right)+\left(4r^{2}+9\right)\mathcal{B}\right)\right)\Omega_{m}\mathcal{H}^{2}}{2a^{2}\beta_{1}\left(r^{2}+1\right)^{2}\mathcal{B}+k^{2}\left(2r^{3}\mathcal{B}+3\beta_{1}\left(r^{2}-1\right)r+3r\mathcal{B}\right)}\delta,\label{eq:solQS_Phi-gen-1}\\
k^{2}\Psi_{f} & =\frac{\left(-3a^{2}\beta_{1}\mathcal{B}\left(9\beta_{1}\left(r^{2}-1\right)r^{2}+\left(r^{2}-2\right)\mathcal{B}\right)-k^{2}r\mathcal{B}\left(9\beta_{1}\left(r^{2}-1\right)+5\mathcal{B}\right)\right)\Omega_{m}\mathcal{H}^{2}}{2a^{2}\beta_{1}\left(r^{2}+1\right)^{2}\mathcal{B}\left(9\beta_{1}\left(r^{2}-1\right)+\mathcal{B}\right)+k^{2}r\left(3\beta_{1}\left(r^{2}-1\right)+\left(2r^{2}+3\right)\mathcal{B}\right)\left(9\beta_{1}\left(r^{2}-1\right)+\mathcal{B}\right)}\delta,\label{eq:solQS_Psif-gen-1}\\
k^{2}\Phi_{f} & =\frac{\left(3a^{2}\beta_{1}\left(r^{2}+1\right)\mathcal{B}+k^{2}r\mathcal{B}\right)\Omega_{m}\mathcal{H}^{2}}{2a^{2}\beta_{1}\left(r^{2}+1\right)^{2}\mathcal{B}+k^{2}\left(2r^{3}\mathcal{B}+3\beta_{1}\left(r^{2}-1\right)r+3r\mathcal{B}\right)}\delta,\label{eq:solQS_Phif-gen-1}\\
k^{2}\Delta E & =\frac{\left(2k^{2}r^{2}\mathcal{B}-\frac{9}{2}a^{2}\beta_{1}r\left(3\beta_{1}\left(r^{2}-1\right)+\mathcal{B}\right)\right)\Omega_{m}\mathcal{H}^{2}}{2a^{4}\beta_{1}^{2}\left(r^{2}+1\right)^{2}\mathcal{B}+\beta_{1}a^{2}k^{2}r\left(3\beta_{1}\left(r^{2}-1\right)+\left(2r^{2}+3\right)\mathcal{B}\right)}\delta,\label{eq:solQS_DeltaE-gen-1}
\end{align}
where we have used the combination $\mathcal{B}\equiv3\beta_{1}\left(r^{2}+1\right)-2B_{1}$
to further simplify the expressions.

In order to compare with observations, we calculate two common modified
gravity parameters: the anisotropic stress, $\eta\equiv-\Phi/\Psi$,
and the effective gravitational coupling for the growth of structures,
$Y\equiv-2k^{2}\Psi/(3\mathcal{H}^{2}\Omega_{m}\delta_{m})$. In general
relativity with $\Lambda$CDM, $\eta=Y=1$, while in $\beta_{1}\beta_{4}$
IBB they possess the following structure, 
\begin{align}
\eta & =H_{2}\frac{1+H_{4}(k/\mathcal{H})^{2}}{1+H_{3}(k/\mathcal{H})^{2}},\label{eq:eta}\\
Y & =H_{1}\frac{1+H_{3}(k/\mathcal{H})^{2}}{1+H_{5}(k/\mathcal{H})^{2}},\label{eq:Y}
\end{align}
with coefficients 
\begin{align}
H_{1} & =-\frac{9\beta_{1}\left(r^{2}-1\right)r^{2}+\left(r^{2}-2\right)\mathcal{B}}{2\left(r^{2}+1\right)^{2}\mathcal{B}},\label{eq:H1}\\
H_{2} & =-\frac{2\left(r^{2}+1\right)\mathcal{B}}{9\beta_{1}\left(r^{2}-1\right)r^{2}+\left(r^{2}-2\right)\mathcal{B}},\\
H_{3} & =-\frac{\mathcal{H}^{2}r\left(9\beta_{1}\left(r^{2}-1\right)+\left(8r^{2}+9\right)\mathcal{B}\right)}{3a^{2}\beta_{1}\left(9\beta_{1}\left(r^{2}-1\right)r^{2}+\left(r^{2}-2\right)\mathcal{B}\right)},\\
H_{4} & =\frac{\mathcal{H}^{2}r\left(9\beta_{1}\left(r^{2}-1\right)+\left(4r^{2}+9\right)\mathcal{B}\right)}{6a^{2}\beta_{1}\left(r^{2}+1\right)\mathcal{B}},\\
H_{5} & =\frac{\mathcal{H}^{2}r\left(6r^{2}\mathcal{B}+9\beta_{1}\left(r^{2}-1\right)+9\mathcal{B}\right)}{6a^{2}\beta_{1}\left(r^{2}+1\right)^{2}\mathcal{B}}.\label{eq:H5}
\end{align}

As a side remark, we note that in this model the asymptotic past corresponds
to the limit $r\rightarrow\infty$ and $r'\rightarrow-\frac{3}{2}r$,
i.e., $r\rightarrow a^{-3/2}$. This implies that $b\sim a^{-1/2}$,
i.e., the second metric initially collapses while ``our'' metric
expands. On the approach to the final de Sitter stage, $r$ approaches
a constant $r_{c}$, so the scale factors $a$ and $b$ both expand
exponentially. The $f$-metric scale factor, $b$, therefore undergoes
a bounce in this model.

This bounce has an unusual consequence. Recall from Eq.~(\ref{metric_f}) that, after imposing the Bianchi identity, we have $f_{00} = -\dot{b}^2/\mathcal{H}^2$. Therefore, when $b$ bounces, $f_{00}$ becomes zero: at that one point, the lapse function of the
$f$ metric vanishes.\footnote{Moreover, the square root of this, $\dot b/\mathcal{H}$, appears in the mass terms. This quantity starts off negative at early times and then becomes positive.} We believe, however, that this does not render the solution
unphysical, for the following reasons. First, the $f$ metric does not couple
to matter and so, unlike the $g$ metric, it does not have a geometric interpretation.
A singularity in the $f$-metric therefore does not necessarily imply a
singularity in  observable quantities. In fact, we find no singularity in
any of our background or perturbed variables. Second, although the Riemann
tensor for the $f$ metric is singular when $f_{00}=0$, the Lagrangian
density $\sqrt{-\det f} R_f$ remains finite and nonzero at all times, so
the equations of motion can be derived at any points in time.

\begin{figure}
\includegraphics[width=0.6\textwidth]{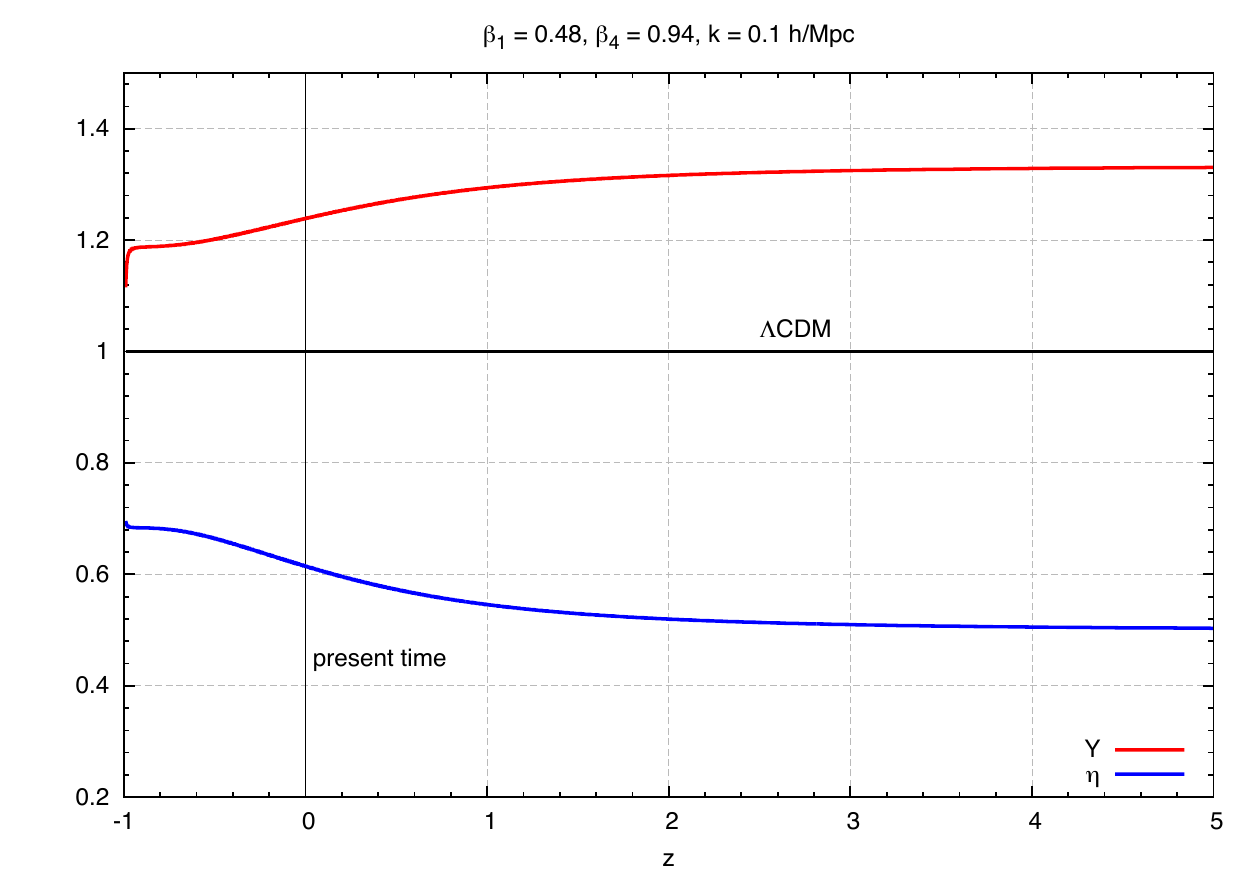} \protect\caption{The modified-gravity parameters, $Y$ and $\eta$, for the $\beta_{1}\beta_{4}$ IBB model, from $z=5$ until the asymptotic (de Sitter) future. Notice
that the parameters approach a constant late-time value until a late
era of horizon exit, when the $k=0.1~h$/Mpc mode becomes superhorizon
and the QS limit breaks down. The horizontal line corresponds to the $\Lambda$CDM prediction for $Y$ and $\eta$, and the vertical line is the present day. These curves are very weakly dependent on $k$. For concreteness, we use the best-fit values $\beta_1=0.48$ and $\beta_4=0.94$, calculated in Sec.~\ref{sec:comparison}.}

\label{fig:b1b4mg} 
\end{figure}

In the asymptotic past, every infinite-branch $\beta_{1}\beta_{4}$
model satisfies 
\begin{equation}
\lim_{N\rightarrow-\infty}\eta=\frac{1}{2}\qquad\text{and}\qquad\lim_{N\rightarrow-\infty}Y=\frac{4}{3}
\end{equation}
and therefore does not reduce to the standard $\Lambda$CDM. In the
future one finds $\eta\to1$ if $k$ is kept finite, but this is somewhat
fictitious: for any finite $k$ there will be an epoch of horizon
exit in the future after which the subhorizon QS approximation breaks down. We
can see both this asymptotic past and future behavior in Fig.~\ref{fig:b1b4mg},
although the late-time approach of $\eta$ to unity is not easily
visible.

\section{Comparison to measured growth data}
\label{sec:comparison}

In this section we compare the predictions in the quasi-static approximation to
the measured growth rate. In Ref.~\cite{2014arXiv1404.4061S}, we discussed the
numerical results of the modified-gravity parameters, Eq. (\ref{eq:eta},
\ref{eq:Y}), for $\beta_{1}\beta_{4}$ infinite-branch bigravity and their early time limits,\footnote{Note that Ref.~\cite{2014arXiv1404.4061S} uses a slightly different
effective gravitational constant, $Q\equiv\eta Y$.%
} and compared to the data. Although we found strong deviations
from the $\Lambda$CDM values, the model is at present still in agreement
with the observed growth data. However, as we mentioned,
future experiments will be able to distinguish between the predictions
of the $\Lambda$CDM and bimetric gravity for $\eta$ and $Y$.

We use the data set compiled by Ref.~\cite{Macaulay:2013swa} containing
the current measurements of the quantity 
\begin{equation}
f(z)\sigma_{8}(z)=f(z)G(z)\sigma_{8},
\end{equation}
where $f(z)\equiv\delta'/\delta$ and $G(z)$ is
the growth factor normalized to the present. The data come from the
6dFGS \cite{Beutler:2012px}, LRG$_{200}$, LRG$_{60}$ \cite{Samushia:2011cs},
BOSS \cite{Tojeiro:2012rp}, WiggleZ \cite{Blake:2012pj}, and VIPERS
\cite{delaTorre:2013rpa} surveys. These measurements
can be compared to the theoretical growth rate which follows from
integrating Eq. (\ref{eq:delta_full}) in the QS limit: 
\begin{equation}
\delta_{m}''+\delta_{m}'\left(1+\frac{\mathcal{H}'}{\mathcal{H}}\right)-\frac{3}{2}Y(k)\Omega_{m}\delta_{m}=0.
\end{equation}
The theoretically expected and observed data, $t_{i}$ and $d_{i}$,
respectively, can be compared to compute 
\begin{align}
\chi_{f\sigma_{8}}^{2} & =\sum_{ij}\left(d_{i}-\sigma_{8}t_{i}\right)C_{ij}^{-1}\left(d_{j}-\sigma_{8}t_{j}\right),
\end{align}
where $C_{ij}$ denotes the covariance matrix. Since no model-free
constraints on $\sigma_{8}$ exist, one can remove this dependency
with a marginalization over positive values which can be performed
analytically: 
\begin{equation}
\chi_{f\sigma_{8}}^{2}=S_{20}-\frac{S_{11}^{2}}{S_{02}}+\log S_{02}-2\log\left(1+\mathrm{Erf}\left(\frac{S_{11}}{\sqrt{2S_{02}}}\right)\right).\label{eq:chimarg}
\end{equation}
Here, $S_{11}=d_{i}C_{ij}^{-1}t_{j},\,$ $S_{20}=d_{i}C_{ij}^{-1}d_{j},$
and $S_{02}=t_{i}C_{ij}^{-1}t_{j}.$ Note that $Y$ is (weakly) scale-dependent
but the current observational data are averaged over a range of scales.
For the computation of the likelihood, we assume an average scale
$k=0.1h/\text{Mpc}$.

\begin{figure}
\includegraphics[width=0.8\textwidth]{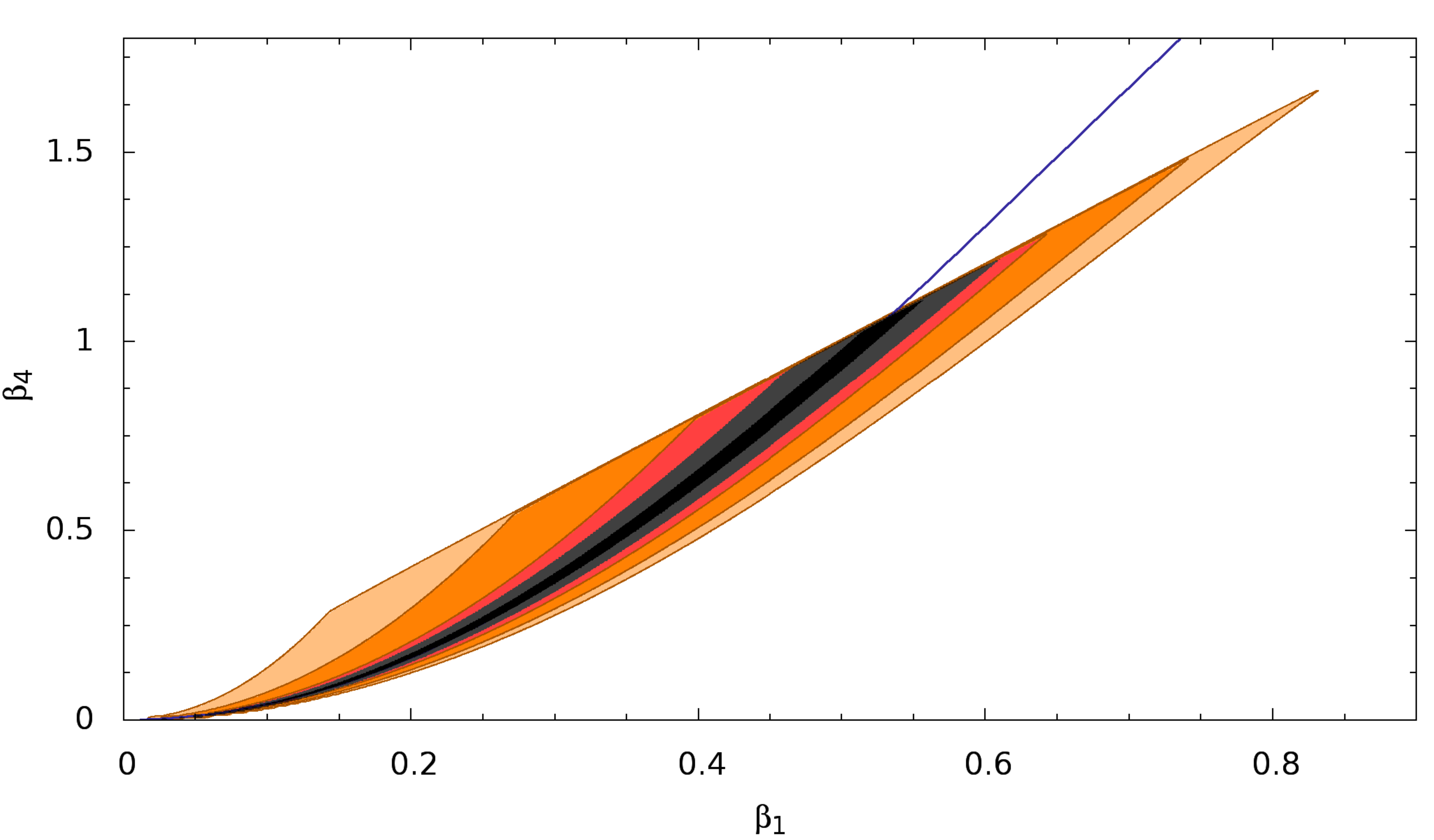}

\protect\caption{Likelihood from measured growth rates, where the red, orange, and light
orange filled regions correspond to 68\%, 95\% and 99.7\% confidence
levels. Both black (68\%) and gray (99.7\%) regions illustrate the
combination of the likelihoods from measured growth data and type Ia supernovae.
The blue line indicates the degeneracy curve corresponding to the
background best-fit points. Note that the viability condition enforces
the likelihood to vanish when $\beta_{4}>2\beta_{2}$.}
\label{fig:likelihood}
\end{figure}

As shown in Fig. \ref{fig:likelihood}, the
confidence region obtained from the growth data is in agreement with
type Ia SNe data (see Ref.~\cite{1475-7516-2014-03-029} for the likelihood
from the SCP Union 2.1 Compilation of SNe Ia data \cite{Suzuki:2011hu}).
The growth data alone provides $\beta_{1}=0.40_{-0.15}^{+0.14}$ and
$\beta_{4}=0.67_{-0.38}^{+0.31}$ with a $\chi_{\mathrm{min}}^{2}=9.72$ (with nine degrees of freedom)
for the best-fit value and is in agreement with the SNe Ia likelihood.
The likelihood from growth data is, however, a much weaker constraint than the likelihood from background observations.
Thus, the combination of both likelihoods, providing $\beta_{1}=0.48_{-0.16}^{+0.05}$
and $\beta_{4}=0.94_{-0.51}^{+0.11}$, is similar to the SNe Ia result
alone.

\begin{figure}

\includegraphics[width=0.6\textwidth]{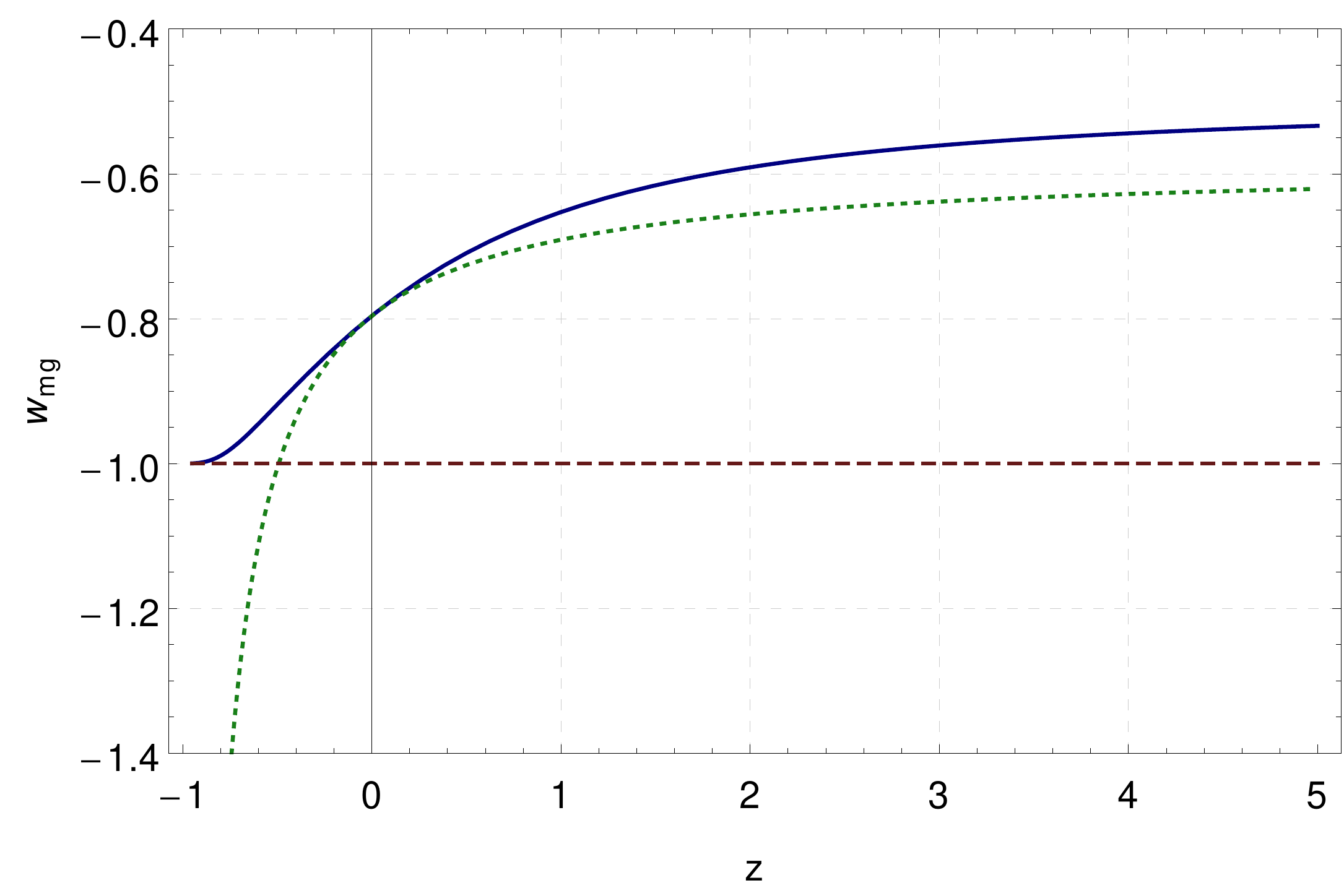}\protect\caption{The equation of state (EOS, solid blue) in the IBB model with $\beta_{1}=0.48$, $\beta_{4}=0.94$, along with the CPL approximation $w(a) \approx w_0 + w_az/(1+z)$ (dotted green) where $w_a$ corresponds to the slope at present time. In the asymptotic future,
$w_{mg}$ tends to $-1$, i.e., the EOS of a cosmological constant (dashed red).}
\label{fig:EOS}

\end{figure}

Note that those favored parameter regions
were obtained by integrating the two-dimensional likelihood and are
not Gaussian distributed due to the degeneracy in the parameters $\beta_{1}$
and $\beta_{4}$ (see Eq. (\ref{eq:degeneracy_curve_b1b4})). This degeneracy
curve is unaffected by additional growth data and is still parametrized
by the SNe Ia result $\Omega_{m0}=1-\Omega_{mg0}=0.16_{-0.03}^{+0.02}$
(note that the combination of the most likely parameters predicts,
however, $\Omega_{m0}=0.18$). According to Eq. (\ref{eq:w_mg}), the
EOS of modified gravity, $w_\mathrm{mg}$, is best fit by $w_{0}=-0.79$ and $w_{a}=0.21$, where we use
the Chevallier-Polarski-Linder (CPL) parametrization \cite{2001IJMPD..10..213C,2003PhRvL..90i1301L},
\begin{equation}
w(z)= w_0 + w_az/(1+z).\label{eq:CPL}
\end{equation}
However, since we approximated the EOS near the present time, we cannot expect Eq. (\ref{eq:CPL}) to fit the real EOS well at early
times or in the future. As shown in Fig.~\ref{fig:EOS}, the fit is in fact valid in the past only up to $z\approx 0.5$, while in the future the limit $w_{mg}\to-1$ is lost.

\begin{figure}
\includegraphics[width=0.8\textwidth]{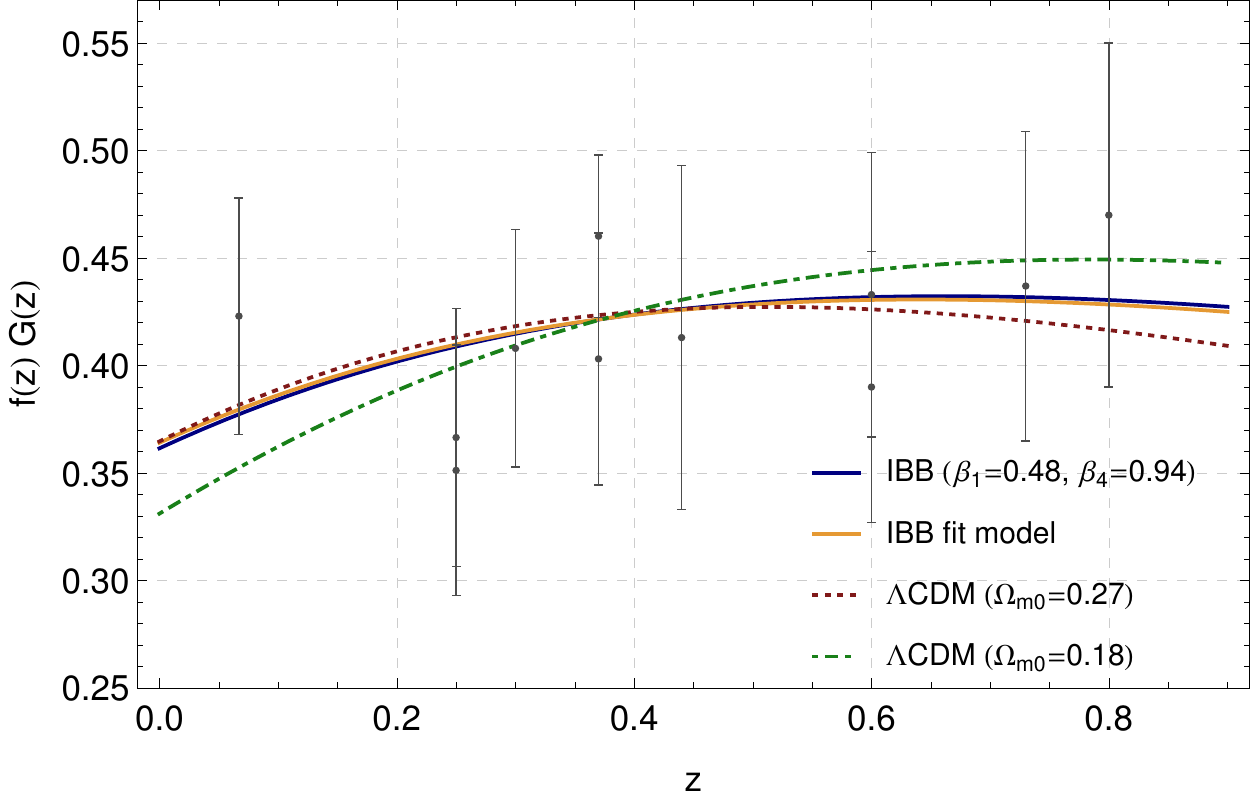}

\protect\caption{Growth history for the best-fit IBB model (solid blue)
with $\beta_{1}=0.48$ and $\beta_{4}=0.94$ compared to the result obtained from the best fit (\ref{eq:fitmodel_f}) (solid orange) with $\gamma_{0}=0.47$ and $\alpha=0.21$, and the $\Lambda$CDM
predictions for $\Omega_{m0}=0.27$ (dotted red) and $\Omega_{m0}=0.18$
(dotted-dashed green). The latter value for the matter density is
similar to that corresponding to IBB.
Note that a vertical shift of each single curve is possible due to
the marginalization over $\sigma_{8}$. Here, we choose that value
for $\sigma_{8}$ for each curve individually such that it fits the
data best. The growth histories are compared to observed data compiled
by Ref.~\cite{Macaulay:2013swa}.
\label{fig:growthhistory}}
\end{figure}

For one specific choice of parameters, corresponding to the best-fit values, we compared the quantity $f(z)G(z)$ with the measured
growth data and fits from $\Lambda$CDM in Fig. \ref{fig:growthhistory}.
Although the modified-gravity parameters differ significantly from
the $\Lambda$CDM result $Y=\eta=1$, the prediction for $f(z)G(z)$ is in good
agreement with measurements and is close to the $\Lambda$CDM result.

The difference between the growth
rate $f(z)$ in the best-fit model and $\Lambda$CDM is, however,
quite large. Therefore, the common approximation $f\approx\Omega_{m}^{\gamma}$
fits the growth rate very badly, even if the range in the redshift
is small (where $f(z)$ is still smaller than unity) \cite{2014arXiv1404.4061S}.
We have found a two-parameter scheme,
\begin{equation}
f\approx\Omega_{m}^{\gamma_{0}}\left(1+\alpha\frac{z}{1+z}\right),\label{eq:fitmodel_f}
\end{equation}
which is able to provide a much better fit (see Fig. \ref{fig:growthhistory}). Using this approximation,
we obtain $\gamma_{0}=0.47$ and $\alpha=0.21$ as best-fit values.

\section{Conclusions and outlook}

We have investigated the stability of linear cosmological perturbations
in bimetric gravity. Many models with viable
background cosmologies exhibit an instability on small scales until
fairly recently in cosmic history. However, we also found a class
of viable models which are stable at all times: IBB with the interaction parameters $\beta_1$ and $\beta_4$ turned on. In these models, the ratio $r=b/a$ of the
two scale factors decreases from infinity to a finite late-time value.
IBB is able to fit observations at the level of both the
background (type Ia supernovae) and linear, subhorizon perturbations (growth histories)
without requiring an explicit cosmological constant for the physical
metric, although the region of likely parameters is small. The combination
of both likelihoods yields the parameter constraints $\beta_{1}=0.48_{-0.16}^{+0.05}$
and $\beta_{4}=0.94_{-0.51}^{+0.11}$. IBB with these best-fit parameters predicts $\Omega_{m0}=0.18$
and an equation of state $w(z)\approx-0.79+0.21z/(1+z)$. The growth rate, $f \equiv d\ln\delta/d\ln a$, is approximated very well by the two-parameter fit $f(z)\approx\Omega_{m}^{0.47}[1+0.21z/(1+z)]$. Additionally, the two main
modified-gravity parameters, the anisotropic stress $\eta$ and modification
to Newton's constant $Y$, tend to $\eta=\frac{1}{2}$ and $Y=\frac{4}{3}$
for early times and therefore do not reduce to the standard $\Lambda$CDM
result. The predictions of this two-parameter model will be testable
by near-future experiments \cite{Amendola:2013qna}.

On the surface, our results would seem to place in jeopardy a large
swath of bigravity's parameter space, such as the ``minimal'' $\beta_{1}$-only
model which is the only single-parameter model that is viable at the
background level \cite{1475-7516-2014-03-029}. It is important to emphasize that the existence
of such an instability does \textit{not} automatically rule these
models out. It merely impedes our ability to use linear theory on
deep subhorizon scales (recall that the instability is problematic
specifically for large $k$). Models that are not linearly stable
can still be realistic if only the gravitational potentials become
nonlinear, or even if the matter fluctuations also become nonlinear
but in such a way that their properties do not contradict observations.
The theory can be saved if, for instance, the instability is softened
or vanishes entirely when nonlinear effects are taken into account.
We might even expect such behavior: bigravity models exhibit a Vainshtein
mechanism \cite{Vainshtein:1972sx,Babichev:2013usa} which restores
general relativity in environments where the new degrees of freedom
are highly nonlinear.

Consequently there are two very important questions for future work:
can these unstable models still accurately describe the real Universe,
and if so, how can we perform calculations for structure formation?

Until these questions are answered, the $\beta_{1}\beta_{4}$ infinite-branch
model seems to be the most promising target at the moment for studying
bigravity. Because this instability appears to be absent in
the superhorizon limit, it may also be feasible to test the unstable
models using large-scale modes.

What other escape routes are there? Throughout this analysis we have
assumed that only one of the metrics couples to matter. A possible
way to cure bimetric gravity from instabilities while only allowing
one nonvanishing $\beta$ parameter could be to allow matter to couple
to both metrics \cite{2013JCAP...10..046A,Akrami:2014lja}.
In such a theory, the finite-branch solutions asymptote to a nonzero
value for $r$ in the far past, so these theories may avoid the
instability. This would introduce a new coupling parameter, so if
only one $\beta$ parameter is turned on, there are two free parameters
and such a model is arguably as predictive as the $\beta_{1}\beta_{4}$
model. Unfortunately, this way of double-coupling would introduce a ghost \cite{deRham:2014naa,2014arXiv1408.0487Y,2014arXiv1408.5131N}. However, the authors in Ref.~\cite{deRham:2014naa} proposed a coupling to matter using a new composite metric which is free of the ghost in the decoupling limit.  The cosmological background solutions in bigravity with this type of coupling together with a comparison to observations were studied in \cite{Enander:2014xga} (see also Ref \cite{2014arXiv1409.8300S} for the case of massive gravity). The consequences for linear perturbations will be discussed in
a future work (in preparation).

\begin{acknowledgments}
We thank Antonio de Felice, Jonas Enander, Pedro Ferreira, Emir Gumrukcuoglu, Tomi Koivisto, Martin Kunz, Macarena Lagos, Michele Maggiore, Edvard M\"ortsell, Malin Renneby, Jeremy Sakstein, Ippocratis Saltas, Ignacy Sawicki, and Takahiro Tanaka for many useful discussions. L.A. and F.K. acknowledge support from Deutsche Forschungsgemeinschaft (DFG) through
the TR33 project ``The Dark Universe.'' Y.A. is supported by the European Research Council (ERC)
Starting Grant No. StG2010-257080. M.M. acknowledges funding from the Swiss National
Science Foundation. A.R.S. is supported by the
David Gledhill Research Studentship, Sidney Sussex College, University
of Cambridge; and by the Isaac Newton Fund and Studentships, University
of Cambridge. We acknowledge the use of resources
from the Norwegian national supercomputing facilities, NOTUR. 
\end{acknowledgments}

\bibliography{observables,massive-gravity,amendola}

\begin{thebibliography}{63}%
\makeatletter
\providecommand \@ifxundefined [1]{%
 \@ifx{#1\undefined}
}%
\providecommand \@ifnum [1]{%
 \ifnum #1\expandafter \@firstoftwo
 \else \expandafter \@secondoftwo
 \fi
}%
\providecommand \@ifx [1]{%
 \ifx #1\expandafter \@firstoftwo
 \else \expandafter \@secondoftwo
 \fi
}%
\providecommand \natexlab [1]{#1}%
\providecommand \enquote  [1]{``#1''}%
\providecommand \bibnamefont  [1]{#1}%
\providecommand \bibfnamefont [1]{#1}%
\providecommand \citenamefont [1]{#1}%
\providecommand \href@noop [0]{\@secondoftwo}%
\providecommand \href [0]{\begingroup \@sanitize@url \@href}%
\providecommand \@href[1]{\@@startlink{#1}\@@href}%
\providecommand \@@href[1]{\endgroup#1\@@endlink}%
\providecommand \@sanitize@url [0]{\catcode `\\12\catcode `\$12\catcode
  `\&12\catcode `\#12\catcode `\^12\catcode `\_12\catcode `\%12\relax}%
\providecommand \@@startlink[1]{}%
\providecommand \@@endlink[0]{}%
\providecommand \url  [0]{\begingroup\@sanitize@url \@url }%
\providecommand \@url [1]{\endgroup\@href {#1}{\urlprefix }}%
\providecommand \urlprefix  [0]{URL }%
\providecommand \Eprint [0]{\href }%
\providecommand \doibase [0]{http://dx.doi.org/}%
\providecommand \selectlanguage [0]{\@gobble}%
\providecommand \bibinfo  [0]{\@secondoftwo}%
\providecommand \bibfield  [0]{\@secondoftwo}%
\providecommand \translation [1]{[#1]}%
\providecommand \BibitemOpen [0]{}%
\providecommand \bibitemStop [0]{}%
\providecommand \bibitemNoStop [0]{.\EOS\space}%
\providecommand \EOS [0]{\spacefactor3000\relax}%
\providecommand \BibitemShut  [1]{\csname bibitem#1\endcsname}%
\let\auto@bib@innerbib\@empty
%</preamble>
\bibitem [{\citenamefont {Gupta}(1954)}]{Gupta:1954zz}%
  \BibitemOpen
  \bibfield  {author} {\bibinfo {author} {\bibfnamefont {Suraj~N.}\
  \bibnamefont {Gupta}},\ }\bibfield  {title} {\enquote {\bibinfo {title}
  {{Gravitation and Electromagnetism}},}\ }\href {\doibase
  10.1103/PhysRev.96.1683} {\bibfield  {journal} {\bibinfo  {journal}
  {Phys.Rev.}\ }\textbf {\bibinfo {volume} {96}},\ \bibinfo {pages}
  {1683--1685} (\bibinfo {year} {1954})}\BibitemShut {NoStop}%
%\%CITATION = PHRVA,96,1683;\%\%
\bibitem [{\citenamefont {Weinberg}(1965)}]{Weinberg:1965rz}%
  \BibitemOpen
  \bibfield  {author} {\bibinfo {author} {\bibfnamefont {Steven}\ \bibnamefont
  {Weinberg}},\ }\bibfield  {title} {\enquote {\bibinfo {title} {{Photons and
  gravitons in perturbation theory: Derivation of Maxwell's and Einstein's
  equations}},}\ }\href {\doibase 10.1103/PhysRev.138.B988} {\bibfield
  {journal} {\bibinfo  {journal} {Phys.Rev.}\ }\textbf {\bibinfo {volume}
  {138}},\ \bibinfo {pages} {B988--B1002} (\bibinfo {year} {1965})}\BibitemShut
  {NoStop}%
%\%CITATION = PHRVA,138,B988;\%\%
\bibitem [{\citenamefont {Deser}(1970)}]{Deser:1969wk}%
  \BibitemOpen
  \bibfield  {author} {\bibinfo {author} {\bibfnamefont {Stanley}\ \bibnamefont
  {Deser}},\ }\bibfield  {title} {\enquote {\bibinfo {title} {{Selfinteraction
  and gauge invariance}},}\ }\href {\doibase 10.1007/BF00759198} {\bibfield
  {journal} {\bibinfo  {journal} {Gen.Rel.Grav.}\ }\textbf {\bibinfo {volume}
  {1}},\ \bibinfo {pages} {9--18} (\bibinfo {year} {1970})},\ \Eprint
  {http://arxiv.org/abs/gr-qc/0411023} {arXiv:gr-qc/0411023 [gr-qc]}
  \BibitemShut {NoStop}%
%\%CITATION = GR-QC/0411023;\%\%
\bibitem [{\citenamefont {Boulware}\ and\ \citenamefont
  {Deser}(1975)}]{Boulware:1974sr}%
  \BibitemOpen
  \bibfield  {author} {\bibinfo {author} {\bibfnamefont {David~G.}\
  \bibnamefont {Boulware}}\ and\ \bibinfo {author} {\bibfnamefont {Stanley}\
  \bibnamefont {Deser}},\ }\bibfield  {title} {\enquote {\bibinfo {title}
  {{Classical General Relativity Derived from Quantum Gravity}},}\ }\href
  {\doibase 10.1016/0003-4916(75)90302-4} {\bibfield  {journal} {\bibinfo
  {journal} {Annals Phys.}\ }\textbf {\bibinfo {volume} {89}},\ \bibinfo
  {pages} {193} (\bibinfo {year} {1975})}\BibitemShut {NoStop}%
%\%CITATION = APNYA,89,193;\%\%
\bibitem [{\citenamefont {Feynman}\ \emph {et~al.}(1996)\citenamefont
  {Feynman}, \citenamefont {Morinigo}, \citenamefont {Wagner},\ and\
  \citenamefont {Hatfield}}]{Feynman:1996kb}%
  \BibitemOpen
  \bibfield  {author} {\bibinfo {author} {\bibfnamefont {R.P.}\ \bibnamefont
  {Feynman}}, \bibinfo {author} {\bibfnamefont {F.B.}\ \bibnamefont
  {Morinigo}}, \bibinfo {author} {\bibfnamefont {W.G.}\ \bibnamefont {Wagner}},
  \ and\ \bibinfo {author} {\bibfnamefont {B.}~\bibnamefont {Hatfield}},\
  }\bibfield  {title} {\enquote {\bibinfo {title} {{Feynman lectures on
  gravitation}},}\ }\href@noop {} {\  (\bibinfo {year} {1996})}\BibitemShut
  {NoStop}%
%\%CITATION = INSPIRE-427379;\%\%
\bibitem [{\citenamefont {{Deser}}(2010)}]{2010GReGr..42..641D}%
  \BibitemOpen
  \bibfield  {author} {\bibinfo {author} {\bibfnamefont {S.}~\bibnamefont
  {{Deser}}},\ }\bibfield  {title} {\enquote {\bibinfo {title} {{Gravity from
  self-interaction redux}},}\ }\href {\doibase 10.1007/s10714-009-0912-9}
  {\bibfield  {journal} {\bibinfo  {journal} {General Relativity and
  Gravitation}\ }\textbf {\bibinfo {volume} {42}},\ \bibinfo {pages} {641--646}
  (\bibinfo {year} {2010})},\ \Eprint {http://arxiv.org/abs/0910.2975}
  {arXiv:0910.2975 [gr-qc]} \BibitemShut {NoStop}%
\bibitem [{\citenamefont {{Barcel{\'o}}}\ \emph
  {et~al.}(2014{\natexlab{a}})\citenamefont {{Barcel{\'o}}}, \citenamefont
  {{Carballo-Rubio}},\ and\ \citenamefont {{Garay}}}]{2014PhRvD..89l4019B}%
  \BibitemOpen
  \bibfield  {author} {\bibinfo {author} {\bibfnamefont {C.}~\bibnamefont
  {{Barcel{\'o}}}}, \bibinfo {author} {\bibfnamefont {R.}~\bibnamefont
  {{Carballo-Rubio}}}, \ and\ \bibinfo {author} {\bibfnamefont {L.~J.}\
  \bibnamefont {{Garay}}},\ }\bibfield  {title} {\enquote {\bibinfo {title}
  {{Unimodular gravity and general relativity from graviton
  self-interactions}},}\ }\href {\doibase 10.1103/PhysRevD.89.124019}
  {\bibfield  {journal} {\bibinfo  {journal} {\prd}\ }\textbf {\bibinfo
  {volume} {89}},\ \bibinfo {eid} {124019} (\bibinfo {year}
  {2014}{\natexlab{a}})},\ \Eprint {http://arxiv.org/abs/1401.2941}
  {arXiv:1401.2941 [gr-qc]} \BibitemShut {NoStop}%
\bibitem [{\citenamefont {{Barcel{\'o}}}\ \emph
  {et~al.}(2014{\natexlab{b}})\citenamefont {{Barcel{\'o}}}, \citenamefont
  {{Carballo-Rubio}},\ and\ \citenamefont {{Garay}}}]{2014arXiv1406.7713B}%
  \BibitemOpen
  \bibfield  {author} {\bibinfo {author} {\bibfnamefont {C.}~\bibnamefont
  {{Barcel{\'o}}}}, \bibinfo {author} {\bibfnamefont {R.}~\bibnamefont
  {{Carballo-Rubio}}}, \ and\ \bibinfo {author} {\bibfnamefont {L.~J.}\
  \bibnamefont {{Garay}}},\ }\bibfield  {title} {\enquote {\bibinfo {title}
  {{Absence of cosmological constant problem in special relativistic field
  theory of gravity}},}\ }\href@noop {} {\bibfield  {journal} {\bibinfo
  {journal} {ArXiv e-prints}\ } (\bibinfo {year} {2014}{\natexlab{b}})},\
  \Eprint {http://arxiv.org/abs/1406.7713} {arXiv:1406.7713 [gr-qc]}
  \BibitemShut {NoStop}%
\bibitem [{\citenamefont {Horndeski}(1974)}]{Horndeski:1974}%
  \BibitemOpen
  \bibfield  {author} {\bibinfo {author} {\bibfnamefont {Gregory~Walter}\
  \bibnamefont {Horndeski}},\ }\bibfield  {title} {\enquote {\bibinfo {title}
  {{Second-order scalar-tensor field equations in a four-dimensional space}},}\
  }\href@noop {} {\bibfield  {journal} {\bibinfo  {journal} {Int.J.Th.Phys.}\
  }\textbf {\bibinfo {volume} {10}},\ \bibinfo {pages} {363--384} (\bibinfo
  {year} {1974})}\BibitemShut {NoStop}%
\bibitem [{\citenamefont {Deffayet}\ \emph {et~al.}(2011)\citenamefont
  {Deffayet}, \citenamefont {Gao}, \citenamefont {Steer},\ and\ \citenamefont
  {Zahariade}}]{Deffayet:2011gz}%
  \BibitemOpen
  \bibfield  {author} {\bibinfo {author} {\bibfnamefont {C.}~\bibnamefont
  {Deffayet}}, \bibinfo {author} {\bibfnamefont {Xian}\ \bibnamefont {Gao}},
  \bibinfo {author} {\bibfnamefont {D.A.}\ \bibnamefont {Steer}}, \ and\
  \bibinfo {author} {\bibfnamefont {G.}~\bibnamefont {Zahariade}},\ }\bibfield
  {title} {\enquote {\bibinfo {title} {{From k-essence to generalised
  Galileons}},}\ }\href@noop {} {\bibfield  {journal} {\bibinfo  {journal}
  {Phys.Rev.}\ }\textbf {\bibinfo {volume} {D84}},\ \bibinfo {pages} {064039}
  (\bibinfo {year} {2011})},\ \Eprint {http://arxiv.org/abs/1103.3260}
  {arXiv:1103.3260 [hep-th]} \BibitemShut {NoStop}%
%\%CITATION = ARXIV:1103.3260;\%\%
\bibitem [{\citenamefont {Jacobson}(2007)}]{Jacobson:2008aj}%
  \BibitemOpen
  \bibfield  {author} {\bibinfo {author} {\bibfnamefont {Ted}\ \bibnamefont
  {Jacobson}},\ }\bibfield  {title} {\enquote {\bibinfo {title}
  {{Einstein-aether gravity: A Status report}},}\ }\href@noop {} {\bibfield
  {journal} {\bibinfo  {journal} {PoS}\ }\textbf {\bibinfo {volume} {QG-PH}},\
  \bibinfo {pages} {020} (\bibinfo {year} {2007})},\ \Eprint
  {http://arxiv.org/abs/0801.1547} {arXiv:0801.1547 [gr-qc]} \BibitemShut
  {NoStop}%
%\%CITATION = ARXIV:0801.1547;\%\%
\bibitem [{\citenamefont {Solomon}\ and\ \citenamefont
  {Barrow}(2014)}]{Solomon:2013iza}%
  \BibitemOpen
  \bibfield  {author} {\bibinfo {author} {\bibfnamefont {Adam~R.}\ \bibnamefont
  {Solomon}}\ and\ \bibinfo {author} {\bibfnamefont {John~D.}\ \bibnamefont
  {Barrow}},\ }\bibfield  {title} {\enquote {\bibinfo {title} {{Inflationary
  Instabilities of Einstein-Aether Cosmology}},}\ }\href {\doibase
  10.1103/PhysRevD.89.024001} {\bibfield  {journal} {\bibinfo  {journal}
  {Phys.Rev.}\ }\textbf {\bibinfo {volume} {D89}},\ \bibinfo {pages} {024001}
  (\bibinfo {year} {2014})},\ \Eprint {http://arxiv.org/abs/1309.4778}
  {arXiv:1309.4778 [astro-ph.CO]} \BibitemShut {NoStop}%
%\%CITATION = ARXIV:1309.4778;\%\%
\bibitem [{\citenamefont {Fierz}\ and\ \citenamefont
  {Pauli}(1939)}]{Fierz:1939ix}%
  \BibitemOpen
  \bibfield  {author} {\bibinfo {author} {\bibfnamefont {M.}~\bibnamefont
  {Fierz}}\ and\ \bibinfo {author} {\bibfnamefont {W.}~\bibnamefont {Pauli}},\
  }\bibfield  {title} {\enquote {\bibinfo {title} {{On relativistic wave
  equations for particles of arbitrary spin in an electromagnetic field}},}\
  }\href {\doibase 10.1098/rspa.1939.0140} {\bibfield  {journal} {\bibinfo
  {journal} {Proc.Roy.Soc.Lond.}\ }\textbf {\bibinfo {volume} {A173}},\
  \bibinfo {pages} {211--232} (\bibinfo {year} {1939})}\BibitemShut {NoStop}%
%\%CITATION = PRSLA,A173,211;\%\%
\bibitem [{\citenamefont {{Hinterbichler}}(2012)}]{2012RvMP...84..671H}%
  \BibitemOpen
  \bibfield  {author} {\bibinfo {author} {\bibfnamefont {K.}~\bibnamefont
  {{Hinterbichler}}},\ }\bibfield  {title} {\enquote {\bibinfo {title}
  {{Theoretical aspects of massive gravity}},}\ }\href {\doibase
  10.1103/RevModPhys.84.671} {\bibfield  {journal} {\bibinfo  {journal}
  {Reviews of Modern Physics}\ }\textbf {\bibinfo {volume} {84}},\ \bibinfo
  {pages} {671--710} (\bibinfo {year} {2012})},\ \Eprint
  {http://arxiv.org/abs/1105.3735} {arXiv:1105.3735 [hep-th]} \BibitemShut
  {NoStop}%
\bibitem [{\citenamefont {de~Rham}(2014)}]{deRham:2014zqa}%
  \BibitemOpen
  \bibfield  {author} {\bibinfo {author} {\bibfnamefont {Claudia}\ \bibnamefont
  {de~Rham}},\ }\bibfield  {title} {\enquote {\bibinfo {title} {{Massive
  Gravity}},}\ }\href@noop {} {\  (\bibinfo {year} {2014})},\ \Eprint
  {http://arxiv.org/abs/1401.4173} {arXiv:1401.4173 [hep-th]} \BibitemShut
  {NoStop}%
%\%CITATION = ARXIV:1401.4173;\%\%
\bibitem [{\citenamefont {de~Rham}\ \emph {et~al.}(2011)\citenamefont
  {de~Rham}, \citenamefont {Gabadadze},\ and\ \citenamefont
  {Tolley}}]{deRham:2010kj}%
  \BibitemOpen
  \bibfield  {author} {\bibinfo {author} {\bibfnamefont {Claudia}\ \bibnamefont
  {de~Rham}}, \bibinfo {author} {\bibfnamefont {Gregory}\ \bibnamefont
  {Gabadadze}}, \ and\ \bibinfo {author} {\bibfnamefont {Andrew~J.}\
  \bibnamefont {Tolley}},\ }\bibfield  {title} {\enquote {\bibinfo {title}
  {{Resummation of Massive Gravity}},}\ }\href {\doibase
  10.1103/PhysRevLett.106.231101} {\bibfield  {journal} {\bibinfo  {journal}
  {Phys.Rev.Lett.}\ }\textbf {\bibinfo {volume} {106}},\ \bibinfo {pages}
  {231101} (\bibinfo {year} {2011})},\ \Eprint {http://arxiv.org/abs/1011.1232}
  {arXiv:1011.1232 [hep-th]} \BibitemShut {NoStop}%
%\%CITATION = ARXIV:1011.1232;\%\%
\bibitem [{\citenamefont {{de Rham}}\ and\ \citenamefont
  {{Gabadadze}}(2010{\natexlab{a}})}]{2010PhRvD..82d4020D}%
  \BibitemOpen
  \bibfield  {author} {\bibinfo {author} {\bibfnamefont {C.}~\bibnamefont {{de
  Rham}}}\ and\ \bibinfo {author} {\bibfnamefont {G.}~\bibnamefont
  {{Gabadadze}}},\ }\bibfield  {title} {\enquote {\bibinfo {title}
  {{Generalization of the Fierz-Pauli action}},}\ }\href {\doibase
  10.1103/PhysRevD.82.044020} {\bibfield  {journal} {\bibinfo  {journal}
  {\prd}\ }\textbf {\bibinfo {volume} {82}},\ \bibinfo {eid} {044020} (\bibinfo
  {year} {2010}{\natexlab{a}})},\ \Eprint {http://arxiv.org/abs/1007.0443}
  {arXiv:1007.0443 [hep-th]} \BibitemShut {NoStop}%
\bibitem [{\citenamefont {{de Rham}}\ and\ \citenamefont
  {{Gabadadze}}(2010{\natexlab{b}})}]{2010PhLB..693..334D}%
  \BibitemOpen
  \bibfield  {author} {\bibinfo {author} {\bibfnamefont {C.}~\bibnamefont {{de
  Rham}}}\ and\ \bibinfo {author} {\bibfnamefont {G.}~\bibnamefont
  {{Gabadadze}}},\ }\bibfield  {title} {\enquote {\bibinfo {title} {{Selftuned
  massive spin-2}},}\ }\href {\doibase 10.1016/j.physletb.2010.08.043}
  {\bibfield  {journal} {\bibinfo  {journal} {Physics Letters B}\ }\textbf
  {\bibinfo {volume} {693}},\ \bibinfo {pages} {334--338} (\bibinfo {year}
  {2010}{\natexlab{b}})},\ \Eprint {http://arxiv.org/abs/1006.4367}
  {arXiv:1006.4367 [hep-th]} \BibitemShut {NoStop}%
\bibitem [{\citenamefont {Hassan}\ and\ \citenamefont
  {Rosen}(2011)}]{Hassan:2011vm}%
  \BibitemOpen
  \bibfield  {author} {\bibinfo {author} {\bibfnamefont {S.F.}\ \bibnamefont
  {Hassan}}\ and\ \bibinfo {author} {\bibfnamefont {Rachel~A.}\ \bibnamefont
  {Rosen}},\ }\bibfield  {title} {\enquote {\bibinfo {title} {{On Non-Linear
  Actions for Massive Gravity}},}\ }\href {\doibase 10.1007/JHEP07(2011)009}
  {\bibfield  {journal} {\bibinfo  {journal} {JHEP}\ }\textbf {\bibinfo
  {volume} {1107}},\ \bibinfo {pages} {009} (\bibinfo {year} {2011})},\ \Eprint
  {http://arxiv.org/abs/1103.6055} {arXiv:1103.6055 [hep-th]} \BibitemShut
  {NoStop}%
%\%CITATION = ARXIV:1103.6055;\%\%
\bibitem [{\citenamefont {Hassan}\ and\ \citenamefont
  {Rosen}(2012)}]{Hassan:2011hr}%
  \BibitemOpen
  \bibfield  {author} {\bibinfo {author} {\bibfnamefont {S.F.}\ \bibnamefont
  {Hassan}}\ and\ \bibinfo {author} {\bibfnamefont {Rachel~A.}\ \bibnamefont
  {Rosen}},\ }\bibfield  {title} {\enquote {\bibinfo {title} {{Resolving the
  Ghost Problem in non-Linear Massive Gravity}},}\ }\href {\doibase
  10.1103/PhysRevLett.108.041101} {\bibfield  {journal} {\bibinfo  {journal}
  {Phys.Rev.Lett.}\ }\textbf {\bibinfo {volume} {108}},\ \bibinfo {pages}
  {041101} (\bibinfo {year} {2012})},\ \Eprint {http://arxiv.org/abs/1106.3344}
  {arXiv:1106.3344 [hep-th]} \BibitemShut {NoStop}%
%\%CITATION = ARXIV:1106.3344;\%\%
\bibitem [{\citenamefont {Maggiore}\ and\ \citenamefont
  {Mancarella}(2014)}]{Maggiore:2014sia}%
  \BibitemOpen
  \bibfield  {author} {\bibinfo {author} {\bibfnamefont {Michele}\ \bibnamefont
  {Maggiore}}\ and\ \bibinfo {author} {\bibfnamefont {Michele}\ \bibnamefont
  {Mancarella}},\ }\bibfield  {title} {\enquote {\bibinfo {title} {{Non-local
  gravity and dark energy}},}\ }\href@noop {} {\  (\bibinfo {year} {2014})},\
  \Eprint {http://arxiv.org/abs/1402.0448} {arXiv:1402.0448 [hep-th]}
  \BibitemShut {NoStop}%
%\%CITATION = ARXIV:1402.0448;\%\%
\bibitem [{\citenamefont {{Hassan}}\ \emph {et~al.}(2012)\citenamefont
  {{Hassan}}, \citenamefont {{Rosen}},\ and\ \citenamefont
  {{Schmidt-May}}}]{2012JHEP...02..026H}%
  \BibitemOpen
  \bibfield  {author} {\bibinfo {author} {\bibfnamefont {S.~F.}\ \bibnamefont
  {{Hassan}}}, \bibinfo {author} {\bibfnamefont {R.~A.}\ \bibnamefont
  {{Rosen}}}, \ and\ \bibinfo {author} {\bibfnamefont {A.}~\bibnamefont
  {{Schmidt-May}}},\ }\bibfield  {title} {\enquote {\bibinfo {title}
  {{Ghost-free massive gravity with a general reference metric}},}\ }\href
  {\doibase 10.1007/JHEP02(2012)026} {\bibfield  {journal} {\bibinfo  {journal}
  {Journal of High Energy Physics}\ }\textbf {\bibinfo {volume} {2}},\ \bibinfo
  {pages} {26} (\bibinfo {year} {2012})},\ \Eprint
  {http://arxiv.org/abs/1109.3230} {arXiv:1109.3230 [hep-th]} \BibitemShut
  {NoStop}%
\bibitem [{\citenamefont {{Hassan}}\ and\ \citenamefont
  {{Rosen}}(2012)}]{2012JHEP...02..126H}%
  \BibitemOpen
  \bibfield  {author} {\bibinfo {author} {\bibfnamefont {S.~F.}\ \bibnamefont
  {{Hassan}}}\ and\ \bibinfo {author} {\bibfnamefont {R.~A.}\ \bibnamefont
  {{Rosen}}},\ }\bibfield  {title} {\enquote {\bibinfo {title} {{Bimetric
  gravity from ghost-free massive gravity}},}\ }\href {\doibase
  10.1007/JHEP02(2012)126} {\bibfield  {journal} {\bibinfo  {journal} {Journal
  of High Energy Physics}\ }\textbf {\bibinfo {volume} {2}},\ \bibinfo {pages}
  {126} (\bibinfo {year} {2012})},\ \Eprint {http://arxiv.org/abs/1109.3515}
  {arXiv:1109.3515 [hep-th]} \BibitemShut {NoStop}%
\bibitem [{\citenamefont {von Strauss}\ \emph {et~al.}(2012)\citenamefont {von
  Strauss}, \citenamefont {Schmidt-May}, \citenamefont {Enander}, \citenamefont
  {Mortsell},\ and\ \citenamefont {Hassan}}]{2012JCAP...03..042V}%
  \BibitemOpen
  \bibfield  {author} {\bibinfo {author} {\bibfnamefont {Mikael}\ \bibnamefont
  {von Strauss}}, \bibinfo {author} {\bibfnamefont {Angnis}\ \bibnamefont
  {Schmidt-May}}, \bibinfo {author} {\bibfnamefont {Jonas}\ \bibnamefont
  {Enander}}, \bibinfo {author} {\bibfnamefont {Edvard}\ \bibnamefont
  {Mortsell}}, \ and\ \bibinfo {author} {\bibfnamefont {S.F.}\ \bibnamefont
  {Hassan}},\ }\bibfield  {title} {\enquote {\bibinfo {title} {{Cosmological
  Solutions in Bimetric Gravity and their Observational Tests}},}\ }\href
  {\doibase 10.1088/1475-7516/2012/03/042} {\bibfield  {journal} {\bibinfo
  {journal} {JCAP}\ }\textbf {\bibinfo {volume} {1203}},\ \bibinfo {pages}
  {042} (\bibinfo {year} {2012})},\ \Eprint {http://arxiv.org/abs/1111.1655}
  {arXiv:1111.1655 [gr-qc]} \BibitemShut {NoStop}%
%\%CITATION = ARXIV:1111.1655;\%\%
\bibitem [{\citenamefont {{Akrami}}\ \emph
  {et~al.}(2013{\natexlab{a}})\citenamefont {{Akrami}}, \citenamefont
  {{Koivisto}},\ and\ \citenamefont {{Sandstad}}}]{2013JHEP...03..099A}%
  \BibitemOpen
  \bibfield  {author} {\bibinfo {author} {\bibfnamefont {Y.}~\bibnamefont
  {{Akrami}}}, \bibinfo {author} {\bibfnamefont {T.~S.}\ \bibnamefont
  {{Koivisto}}}, \ and\ \bibinfo {author} {\bibfnamefont {M.}~\bibnamefont
  {{Sandstad}}},\ }\bibfield  {title} {\enquote {\bibinfo {title} {{Accelerated
  expansion from ghost-free bigravity: a statistical analysis with improved
  generality}},}\ }\href {\doibase 10.1007/JHEP03(2013)099} {\bibfield
  {journal} {\bibinfo  {journal} {Journal of High Energy Physics}\ }\textbf
  {\bibinfo {volume} {3}},\ \bibinfo {pages} {99} (\bibinfo {year}
  {2013}{\natexlab{a}})},\ \Eprint {http://arxiv.org/abs/1209.0457}
  {arXiv:1209.0457 [astro-ph.CO]} \BibitemShut {NoStop}%
\bibitem [{\citenamefont {{Akrami}}\ \emph
  {et~al.}(2013{\natexlab{b}})\citenamefont {{Akrami}}, \citenamefont
  {{Koivisto}}, \citenamefont {{Mota}},\ and\ \citenamefont
  {{Sandstad}}}]{2013JCAP...10..046A}%
  \BibitemOpen
  \bibfield  {author} {\bibinfo {author} {\bibfnamefont {Y.}~\bibnamefont
  {{Akrami}}}, \bibinfo {author} {\bibfnamefont {T.~S.}\ \bibnamefont
  {{Koivisto}}}, \bibinfo {author} {\bibfnamefont {D.~F.}\ \bibnamefont
  {{Mota}}}, \ and\ \bibinfo {author} {\bibfnamefont {M.}~\bibnamefont
  {{Sandstad}}},\ }\bibfield  {title} {\enquote {\bibinfo {title} {{Bimetric
  gravity doubly coupled to matter: theory and cosmological implications}},}\
  }\href {\doibase 10.1088/1475-7516/2013/10/046} {\bibfield  {journal}
  {\bibinfo  {journal} {JCAP}\ }\textbf {\bibinfo {volume} {10}},\ \bibinfo
  {eid} {046} (\bibinfo {year} {2013}{\natexlab{b}})},\ \Eprint
  {http://arxiv.org/abs/1306.0004} {arXiv:1306.0004 [hep-th]} \BibitemShut
  {NoStop}%
\bibitem [{\citenamefont {Comelli}\ \emph {et~al.}(2012)\citenamefont
  {Comelli}, \citenamefont {Crisostomi},\ and\ \citenamefont
  {Pilo}}]{Comelli:2012db}%
  \BibitemOpen
  \bibfield  {author} {\bibinfo {author} {\bibfnamefont {D.}~\bibnamefont
  {Comelli}}, \bibinfo {author} {\bibfnamefont {M.}~\bibnamefont {Crisostomi}},
  \ and\ \bibinfo {author} {\bibfnamefont {L.}~\bibnamefont {Pilo}},\
  }\bibfield  {title} {\enquote {\bibinfo {title} {{Perturbations in Massive
  Gravity Cosmology}},}\ }\href {\doibase 10.1007/JHEP06(2012)085} {\bibfield
  {journal} {\bibinfo  {journal} {JHEP}\ }\textbf {\bibinfo {volume} {1206}},\
  \bibinfo {pages} {085} (\bibinfo {year} {2012})},\ \Eprint
  {http://arxiv.org/abs/1202.1986} {arXiv:1202.1986 [hep-th]} \BibitemShut
  {NoStop}%
%\%CITATION = ARXIV:1202.1986;\%\%
\bibitem [{\citenamefont {{De Felice}}\ \emph {et~al.}(2013)\citenamefont {{De
  Felice}}, \citenamefont {{Nakamura}},\ and\ \citenamefont
  {{Tanaka}}}]{2013arXiv1304.3920D}%
  \BibitemOpen
  \bibfield  {author} {\bibinfo {author} {\bibfnamefont {A.}~\bibnamefont {{De
  Felice}}}, \bibinfo {author} {\bibfnamefont {T.}~\bibnamefont {{Nakamura}}},
  \ and\ \bibinfo {author} {\bibfnamefont {T.}~\bibnamefont {{Tanaka}}},\
  }\bibfield  {title} {\enquote {\bibinfo {title} {{Possible existence of
  viable models of bi-gravity with detectable graviton oscillations by
  gravitational wave detectors}},}\ }\href@noop {} {\bibfield  {journal}
  {\bibinfo  {journal} {ArXiv e-prints}\ } (\bibinfo {year} {2013})},\ \Eprint
  {http://arxiv.org/abs/1304.3920} {arXiv:1304.3920 [gr-qc]} \BibitemShut
  {NoStop}%
\bibitem [{\citenamefont {{Comelli}}\ \emph {et~al.}(2012)\citenamefont
  {{Comelli}}, \citenamefont {{Crisostomi}}, \citenamefont {{Nesti}},\ and\
  \citenamefont {{Pilo}}}]{2012JHEP...03..067C}%
  \BibitemOpen
  \bibfield  {author} {\bibinfo {author} {\bibfnamefont {D.}~\bibnamefont
  {{Comelli}}}, \bibinfo {author} {\bibfnamefont {M.}~\bibnamefont
  {{Crisostomi}}}, \bibinfo {author} {\bibfnamefont {F.}~\bibnamefont
  {{Nesti}}}, \ and\ \bibinfo {author} {\bibfnamefont {L.}~\bibnamefont
  {{Pilo}}},\ }\bibfield  {title} {\enquote {\bibinfo {title} {{FRW cosmology
  in ghost free massive gravity from bigravity}},}\ }\href {\doibase
  10.1007/JHEP03(2012)067} {\bibfield  {journal} {\bibinfo  {journal} {Journal
  of High Energy Physics}\ }\textbf {\bibinfo {volume} {3}},\ \bibinfo {pages}
  {67} (\bibinfo {year} {2012})},\ \Eprint {http://arxiv.org/abs/1111.1983}
  {arXiv:1111.1983 [hep-th]} \BibitemShut {NoStop}%
\bibitem [{\citenamefont {{Volkov}}(2012)}]{2012JHEP...01..035V}%
  \BibitemOpen
  \bibfield  {author} {\bibinfo {author} {\bibfnamefont {M.~S.}\ \bibnamefont
  {{Volkov}}},\ }\bibfield  {title} {\enquote {\bibinfo {title} {{Cosmological
  solutions with massive gravitons in the bigravity theory}},}\ }\href
  {\doibase 10.1007/JHEP01(2012)035} {\bibfield  {journal} {\bibinfo  {journal}
  {Journal of High Energy Physics}\ }\textbf {\bibinfo {volume} {1}},\ \bibinfo
  {pages} {35} (\bibinfo {year} {2012})},\ \Eprint
  {http://arxiv.org/abs/1110.6153} {arXiv:1110.6153 [hep-th]} \BibitemShut
  {NoStop}%
\bibitem [{\citenamefont {de~Rham}\ \emph {et~al.}(2013)\citenamefont
  {de~Rham}, \citenamefont {Heisenberg},\ and\ \citenamefont
  {Ribeiro}}]{deRham:2013qqa}%
  \BibitemOpen
  \bibfield  {author} {\bibinfo {author} {\bibfnamefont {Claudia}\ \bibnamefont
  {de~Rham}}, \bibinfo {author} {\bibfnamefont {Lavinia}\ \bibnamefont
  {Heisenberg}}, \ and\ \bibinfo {author} {\bibfnamefont {Raquel~H.}\
  \bibnamefont {Ribeiro}},\ }\bibfield  {title} {\enquote {\bibinfo {title}
  {{Quantum Corrections in Massive Gravity}},}\ }\href {\doibase
  10.1103/PhysRevD.88.084058} {\bibfield  {journal} {\bibinfo  {journal}
  {Phys.Rev.}\ }\textbf {\bibinfo {volume} {D88}},\ \bibinfo {pages} {084058}
  (\bibinfo {year} {2013})},\ \Eprint {http://arxiv.org/abs/1307.7169}
  {arXiv:1307.7169 [hep-th]} \BibitemShut {NoStop}%
%\%CITATION = ARXIV:1307.7169;\%\%
\bibitem [{\citenamefont {Weinberg}(1989)}]{Weinberg:1988cp}%
  \BibitemOpen
  \bibfield  {author} {\bibinfo {author} {\bibfnamefont {Steven}\ \bibnamefont
  {Weinberg}},\ }\bibfield  {title} {\enquote {\bibinfo {title} {{The
  Cosmological Constant Problem}},}\ }\href {\doibase 10.1103/RevModPhys.61.1}
  {\bibfield  {journal} {\bibinfo  {journal} {Rev.Mod.Phys.}\ }\textbf
  {\bibinfo {volume} {61}},\ \bibinfo {pages} {1--23} (\bibinfo {year}
  {1989})}\BibitemShut {NoStop}%
%\%CITATION = RMPHA,61,1;\%\%
\bibitem [{\citenamefont {Martin}(2012)}]{Martin:2012bt}%
  \BibitemOpen
  \bibfield  {author} {\bibinfo {author} {\bibfnamefont {Jerome}\ \bibnamefont
  {Martin}},\ }\bibfield  {title} {\enquote {\bibinfo {title} {{Everything You
  Always Wanted To Know About The Cosmological Constant Problem (But Were
  Afraid To Ask)}},}\ }\href {\doibase 10.1016/j.crhy.2012.04.008} {\bibfield
  {journal} {\bibinfo  {journal} {Comptes Rendus Physique}\ }\textbf {\bibinfo
  {volume} {13}},\ \bibinfo {pages} {566--665} (\bibinfo {year} {2012})},\
  \Eprint {http://arxiv.org/abs/1205.3365} {arXiv:1205.3365 [astro-ph.CO]}
  \BibitemShut {NoStop}%
%\%CITATION = ARXIV:1205.3365;\%\%
\bibitem [{\citenamefont {K{\"o}nnig}\ and\ \citenamefont
  {Amendola}(2014)}]{Konnig:2014dna}%
  \BibitemOpen
  \bibfield  {author} {\bibinfo {author} {\bibfnamefont {Frank}\ \bibnamefont
  {K{\"o}nnig}}\ and\ \bibinfo {author} {\bibfnamefont {Luca}\ \bibnamefont
  {Amendola}},\ }\bibfield  {title} {\enquote {\bibinfo {title} {{Instability
  in a minimal bimetric gravity model}},}\ }\href@noop {} {\  (\bibinfo {year}
  {2014})},\ \Eprint {http://arxiv.org/abs/1402.1988} {arXiv:1402.1988
  [astro-ph.CO]} \BibitemShut {NoStop}%
%\%CITATION = ARXIV:1402.1988;\%\%
\bibitem [{\citenamefont {{Solomon}}\ \emph
  {et~al.}(2014{\natexlab{a}})\citenamefont {{Solomon}}, \citenamefont
  {{Akrami}},\ and\ \citenamefont {{Koivisto}}}]{2014arXiv1404.4061S}%
  \BibitemOpen
  \bibfield  {author} {\bibinfo {author} {\bibfnamefont {A.~R.}\ \bibnamefont
  {{Solomon}}}, \bibinfo {author} {\bibfnamefont {Y.}~\bibnamefont {{Akrami}}},
  \ and\ \bibinfo {author} {\bibfnamefont {T.~S.}\ \bibnamefont {{Koivisto}}},\
  }\bibfield  {title} {\enquote {\bibinfo {title} {{Cosmological perturbations
  in massive bigravity: I. Linear growth of structures}},}\ }\href@noop {}
  {\bibfield  {journal} {\bibinfo  {journal} {ArXiv e-prints}\ } (\bibinfo
  {year} {2014}{\natexlab{a}})},\ \Eprint {http://arxiv.org/abs/1404.4061}
  {arXiv:1404.4061} \BibitemShut {NoStop}%
\bibitem [{\citenamefont {{De Felice}}\ \emph {et~al.}(2014)\citenamefont {{De
  Felice}}, \citenamefont {Gumrukcuoglu}, \citenamefont {Mukohyama},
  \citenamefont {Tanahashi},\ and\ \citenamefont {Tanaka}}]{DeFelice:2014nja}%
  \BibitemOpen
  \bibfield  {author} {\bibinfo {author} {\bibfnamefont {Antonio}\ \bibnamefont
  {{De Felice}}}, \bibinfo {author} {\bibfnamefont {A.~Emir}\ \bibnamefont
  {Gumrukcuoglu}}, \bibinfo {author} {\bibfnamefont {Shinji}\ \bibnamefont
  {Mukohyama}}, \bibinfo {author} {\bibfnamefont {Norihiro}\ \bibnamefont
  {Tanahashi}}, \ and\ \bibinfo {author} {\bibfnamefont {Takahiro}\
  \bibnamefont {Tanaka}},\ }\bibfield  {title} {\enquote {\bibinfo {title}
  {{Viable cosmology in bimetric theory}},}\ }\href@noop {} {\  (\bibinfo
  {year} {2014})},\ \Eprint {http://arxiv.org/abs/1404.0008} {arXiv:1404.0008
  [hep-th]} \BibitemShut {NoStop}%
%\%CITATION = ARXIV:1404.0008;\%\%
\bibitem [{\citenamefont {Comelli}\ \emph {et~al.}(2014)\citenamefont
  {Comelli}, \citenamefont {Crisostomi},\ and\ \citenamefont
  {Pilo}}]{Comelli:2014bqa}%
  \BibitemOpen
  \bibfield  {author} {\bibinfo {author} {\bibfnamefont {D.}~\bibnamefont
  {Comelli}}, \bibinfo {author} {\bibfnamefont {M.}~\bibnamefont {Crisostomi}},
  \ and\ \bibinfo {author} {\bibfnamefont {L.}~\bibnamefont {Pilo}},\
  }\bibfield  {title} {\enquote {\bibinfo {title} {{FRW Cosmological
  Perturbations in Massive Bigravity}},}\ }\href@noop {} {\  (\bibinfo {year}
  {2014})},\ \Eprint {http://arxiv.org/abs/1403.5679} {arXiv:1403.5679
  [hep-th]} \BibitemShut {NoStop}%
%\%CITATION = ARXIV:1403.5679;\%\%
\bibitem [{\citenamefont {{Fasiello}}\ and\ \citenamefont
  {{Tolley}}(2013)}]{2013JCAP...12..002F}%
  \BibitemOpen
  \bibfield  {author} {\bibinfo {author} {\bibfnamefont {M.}~\bibnamefont
  {{Fasiello}}}\ and\ \bibinfo {author} {\bibfnamefont {A.~J.}\ \bibnamefont
  {{Tolley}}},\ }\bibfield  {title} {\enquote {\bibinfo {title} {{Cosmological
  stability bound in massive gravity and bigravity}},}\ }\href {\doibase
  10.1088/1475-7516/2013/12/002} {\bibfield  {journal} {\bibinfo  {journal}
  {JCAP}\ }\textbf {\bibinfo {volume} {12}},\ \bibinfo {eid} {002} (\bibinfo
  {year} {2013})},\ \Eprint {http://arxiv.org/abs/1308.1647} {arXiv:1308.1647
  [hep-th]} \BibitemShut {NoStop}%
\bibitem [{\citenamefont {{Gratia}}\ \emph {et~al.}(2013)\citenamefont
  {{Gratia}}, \citenamefont {{Hu}},\ and\ \citenamefont
  {{Wyman}}}]{2013CQGra..30r4007G}%
  \BibitemOpen
  \bibfield  {author} {\bibinfo {author} {\bibfnamefont {P.}~\bibnamefont
  {{Gratia}}}, \bibinfo {author} {\bibfnamefont {W.}~\bibnamefont {{Hu}}}, \
  and\ \bibinfo {author} {\bibfnamefont {M.}~\bibnamefont {{Wyman}}},\
  }\bibfield  {title} {\enquote {\bibinfo {title} {{Self-accelerating massive
  gravity: how zweibeins walk through determinant singularities}},}\ }\href
  {\doibase 10.1088/0264-9381/30/18/184007} {\bibfield  {journal} {\bibinfo
  {journal} {Classical and Quantum Gravity}\ }\textbf {\bibinfo {volume}
  {30}},\ \bibinfo {eid} {184007} (\bibinfo {year} {2013})},\ \Eprint
  {http://arxiv.org/abs/1305.2916} {arXiv:1305.2916 [hep-th]} \BibitemShut
  {NoStop}%
\bibitem [{\citenamefont {{Gratia}}\ \emph {et~al.}(2014)\citenamefont
  {{Gratia}}, \citenamefont {{Hu}},\ and\ \citenamefont
  {{Wyman}}}]{2014PhRvD..89b7502G}%
  \BibitemOpen
  \bibfield  {author} {\bibinfo {author} {\bibfnamefont {P.}~\bibnamefont
  {{Gratia}}}, \bibinfo {author} {\bibfnamefont {W.}~\bibnamefont {{Hu}}}, \
  and\ \bibinfo {author} {\bibfnamefont {M.}~\bibnamefont {{Wyman}}},\
  }\bibfield  {title} {\enquote {\bibinfo {title} {{Self-accelerating massive
  gravity: Bimetric determinant singularities}},}\ }\href {\doibase
  10.1103/PhysRevD.89.027502} {\bibfield  {journal} {\bibinfo  {journal}
  {\prd}\ }\textbf {\bibinfo {volume} {89}},\ \bibinfo {eid} {027502} (\bibinfo
  {year} {2014})},\ \Eprint {http://arxiv.org/abs/1309.5947} {arXiv:1309.5947
  [hep-th]} \BibitemShut {NoStop}%
\bibitem [{\citenamefont {{Berg}}\ \emph {et~al.}(2012)\citenamefont {{Berg}},
  \citenamefont {{Buchberger}}, \citenamefont {{Enander}}, \citenamefont
  {{M{\"o}rtsell}},\ and\ \citenamefont {{Sj{\"o}rs}}}]{2012JCAP...12..021B}%
  \BibitemOpen
  \bibfield  {author} {\bibinfo {author} {\bibfnamefont {M.}~\bibnamefont
  {{Berg}}}, \bibinfo {author} {\bibfnamefont {I.}~\bibnamefont
  {{Buchberger}}}, \bibinfo {author} {\bibfnamefont {J.}~\bibnamefont
  {{Enander}}}, \bibinfo {author} {\bibfnamefont {E.}~\bibnamefont
  {{M{\"o}rtsell}}}, \ and\ \bibinfo {author} {\bibfnamefont {S.}~\bibnamefont
  {{Sj{\"o}rs}}},\ }\bibfield  {title} {\enquote {\bibinfo {title} {{Growth
  histories in bimetric massive gravity}},}\ }\href {\doibase
  10.1088/1475-7516/2012/12/021} {\bibfield  {journal} {\bibinfo  {journal}
  {JCAP}\ }\textbf {\bibinfo {volume} {12}},\ \bibinfo {eid} {021} (\bibinfo
  {year} {2012})},\ \Eprint {http://arxiv.org/abs/1206.3496} {arXiv:1206.3496
  [gr-qc]} \BibitemShut {NoStop}%
\bibitem [{\citenamefont {Koennig}\ \emph {et~al.}(2014)\citenamefont
  {Koennig}, \citenamefont {Patil},\ and\ \citenamefont
  {Amendola}}]{1475-7516-2014-03-029}%
  \BibitemOpen
  \bibfield  {author} {\bibinfo {author} {\bibfnamefont {Frank}\ \bibnamefont
  {Koennig}}, \bibinfo {author} {\bibfnamefont {Aashay}\ \bibnamefont {Patil}},
  \ and\ \bibinfo {author} {\bibfnamefont {Luca}\ \bibnamefont {Amendola}},\
  }\bibfield  {title} {\enquote {\bibinfo {title} {{Viable cosmological
  solutions in massive bimetric gravity}},}\ }\href {\doibase
  10.1088/1475-7516/2014/03/029} {\bibfield  {journal} {\bibinfo  {journal}
  {Journal of Cosmology and Astroparticle Physics}\ }\textbf {\bibinfo {volume}
  {2014}},\ \bibinfo {pages} {029} (\bibinfo {year} {2014})},\ \Eprint
  {http://arxiv.org/abs/1312.3208} {arXiv:1312.3208 [astro-ph.CO]} \BibitemShut
  {NoStop}%
\bibitem [{\citenamefont {{Khosravi}}\ \emph {et~al.}(2012)\citenamefont
  {{Khosravi}}, \citenamefont {{Sepangi}},\ and\ \citenamefont
  {{Shahidi}}}]{2012PhRvD..86d3517K}%
  \BibitemOpen
  \bibfield  {author} {\bibinfo {author} {\bibfnamefont {N.}~\bibnamefont
  {{Khosravi}}}, \bibinfo {author} {\bibfnamefont {H.~R.}\ \bibnamefont
  {{Sepangi}}}, \ and\ \bibinfo {author} {\bibfnamefont {S.}~\bibnamefont
  {{Shahidi}}},\ }\bibfield  {title} {\enquote {\bibinfo {title} {{Massive
  cosmological scalar perturbations}},}\ }\href {\doibase
  10.1103/PhysRevD.86.043517} {\bibfield  {journal} {\bibinfo  {journal}
  {\prd}\ }\textbf {\bibinfo {volume} {86}},\ \bibinfo {eid} {043517} (\bibinfo
  {year} {2012})},\ \Eprint {http://arxiv.org/abs/1202.2767} {arXiv:1202.2767
  [gr-qc]} \BibitemShut {NoStop}%
\bibitem [{\citenamefont {{Lagos}}\ \emph {et~al.}(2014)\citenamefont
  {{Lagos}}, \citenamefont {{Ba{\~n}ados}}, \citenamefont {{Ferreira}},\ and\
  \citenamefont {{Garc{\'i}a-S{\'a}enz}}}]{2014PhRvD..89b4034L}%
  \BibitemOpen
  \bibfield  {author} {\bibinfo {author} {\bibfnamefont {M.}~\bibnamefont
  {{Lagos}}}, \bibinfo {author} {\bibfnamefont {M.}~\bibnamefont
  {{Ba{\~n}ados}}}, \bibinfo {author} {\bibfnamefont {P.~G.}\ \bibnamefont
  {{Ferreira}}}, \ and\ \bibinfo {author} {\bibfnamefont {S.}~\bibnamefont
  {{Garc{\'i}a-S{\'a}enz}}},\ }\bibfield  {title} {\enquote {\bibinfo {title}
  {{Noether identities and gauge fixing the action for cosmological
  perturbations}},}\ }\href {\doibase 10.1103/PhysRevD.89.024034} {\bibfield
  {journal} {\bibinfo  {journal} {\prd}\ }\textbf {\bibinfo {volume} {89}},\
  \bibinfo {eid} {024034} (\bibinfo {year} {2014})},\ \Eprint
  {http://arxiv.org/abs/1311.3828} {arXiv:1311.3828 [gr-qc]} \BibitemShut
  {NoStop}%
\bibitem [{\citenamefont {Vainshtein}(1972)}]{Vainshtein:1972sx}%
  \BibitemOpen
  \bibfield  {author} {\bibinfo {author} {\bibfnamefont {A.I.}\ \bibnamefont
  {Vainshtein}},\ }\bibfield  {title} {\enquote {\bibinfo {title} {{To the
  problem of nonvanishing gravitation mass}},}\ }\href {\doibase
  10.1016/0370-2693(72)90147-5} {\bibfield  {journal} {\bibinfo  {journal}
  {Phys.Lett.}\ }\textbf {\bibinfo {volume} {B39}},\ \bibinfo {pages}
  {393--394} (\bibinfo {year} {1972})}\BibitemShut {NoStop}%
%\%CITATION = PHLTA,B39,393;\%\%
\bibitem [{\citenamefont {Babichev}\ and\ \citenamefont
  {Deffayet}(2013)}]{Babichev:2013usa}%
  \BibitemOpen
  \bibfield  {author} {\bibinfo {author} {\bibfnamefont {Eugeny}\ \bibnamefont
  {Babichev}}\ and\ \bibinfo {author} {\bibfnamefont {C{\'e}dric}\ \bibnamefont
  {Deffayet}},\ }\bibfield  {title} {\enquote {\bibinfo {title} {{An
  introduction to the Vainshtein mechanism}},}\ }\href {\doibase
  10.1088/0264-9381/30/18/184001} {\bibfield  {journal} {\bibinfo  {journal}
  {Class.Quant.Grav.}\ }\textbf {\bibinfo {volume} {30}},\ \bibinfo {pages}
  {184001} (\bibinfo {year} {2013})},\ \Eprint {http://arxiv.org/abs/1304.7240}
  {arXiv:1304.7240 [gr-qc]} \BibitemShut {NoStop}%
%\%CITATION = ARXIV:1304.7240;\%\%
\bibitem [{\citenamefont {Hassan}\ \emph {et~al.}(2014)\citenamefont {Hassan},
  \citenamefont {Schmidt-May},\ and\ \citenamefont {von
  Strauss}}]{Hassan:2014vja}%
  \BibitemOpen
  \bibfield  {author} {\bibinfo {author} {\bibfnamefont {S.F.}\ \bibnamefont
  {Hassan}}, \bibinfo {author} {\bibfnamefont {Angnis}\ \bibnamefont
  {Schmidt-May}}, \ and\ \bibinfo {author} {\bibfnamefont {Mikael}\
  \bibnamefont {von Strauss}},\ }\bibfield  {title} {\enquote {\bibinfo {title}
  {{Particular Solutions in Bimetric Theory and Their Implications}},}\
  }\href@noop {} {\  (\bibinfo {year} {2014})},\ \Eprint
  {http://arxiv.org/abs/1407.2772} {arXiv:1407.2772 [hep-th]} \BibitemShut
  {NoStop}%
%%CITATION = ARXIV:1407.2772;%%
\bibitem [{\citenamefont {Macaulay}\ \emph {et~al.}(2013)\citenamefont
  {Macaulay}, \citenamefont {Wehus},\ and\ \citenamefont
  {Eriksen}}]{Macaulay:2013swa}%
  \BibitemOpen
  \bibfield  {author} {\bibinfo {author} {\bibfnamefont {Edward}\ \bibnamefont
  {Macaulay}}, \bibinfo {author} {\bibfnamefont {Ingunn~Kathrine}\ \bibnamefont
  {Wehus}}, \ and\ \bibinfo {author} {\bibfnamefont {Hans~Kristian}\
  \bibnamefont {Eriksen}},\ }\bibfield  {title} {\enquote {\bibinfo {title} {{A
  Lower Growth Rate from Recent Redshift Space Distortions than Expected from
  Planck}},}\ }\href@noop {} {\  (\bibinfo {year} {2013})},\ \Eprint
  {http://arxiv.org/abs/1303.6583} {arXiv:1303.6583 [astro-ph.CO]} \BibitemShut
  {NoStop}%
%\%CITATION = ARXIV:1303.6583;\%\%
\bibitem [{\citenamefont {Beutler}\ \emph {et~al.}(2012)\citenamefont
  {Beutler}, \citenamefont {Blake}, \citenamefont {Colless}, \citenamefont
  {Jones}, \citenamefont {Staveley-Smith} \emph {et~al.}}]{Beutler:2012px}%
  \BibitemOpen
  \bibfield  {author} {\bibinfo {author} {\bibfnamefont {Florian}\ \bibnamefont
  {Beutler}}, \bibinfo {author} {\bibfnamefont {Chris}\ \bibnamefont {Blake}},
  \bibinfo {author} {\bibfnamefont {Matthew}\ \bibnamefont {Colless}}, \bibinfo
  {author} {\bibfnamefont {D.~Heath}\ \bibnamefont {Jones}}, \bibinfo {author}
  {\bibfnamefont {Lister}\ \bibnamefont {Staveley-Smith}},  \emph {et~al.},\
  }\bibfield  {title} {\enquote {\bibinfo {title} {{The 6dF Galaxy Survey: z
  approx 0 measurement of the growth rate and sigma8}},}\ }\href@noop {} {\
  (\bibinfo {year} {2012})},\ \Eprint {http://arxiv.org/abs/1204.4725}
  {arXiv:1204.4725 [astro-ph.CO]} \BibitemShut {NoStop}%
%\%CITATION = ARXIV:1204.4725;\%\%
\bibitem [{\citenamefont {Samushia}\ \emph {et~al.}(2012)\citenamefont
  {Samushia}, \citenamefont {Percival},\ and\ \citenamefont
  {Raccanelli}}]{Samushia:2011cs}%
  \BibitemOpen
  \bibfield  {author} {\bibinfo {author} {\bibfnamefont {Lado}\ \bibnamefont
  {Samushia}}, \bibinfo {author} {\bibfnamefont {Will~J.}\ \bibnamefont
  {Percival}}, \ and\ \bibinfo {author} {\bibfnamefont {Alvise}\ \bibnamefont
  {Raccanelli}},\ }\bibfield  {title} {\enquote {\bibinfo {title}
  {{Interpreting large-scale redshift-space distortion measurements}},}\ }\href
  {\doibase 10.1111/j.1365-2966.2011.20169.x} {\bibfield  {journal} {\bibinfo
  {journal} {Mon.Not.Roy.Astron.Soc.}\ }\textbf {\bibinfo {volume} {420}},\
  \bibinfo {pages} {2102--2119} (\bibinfo {year} {2012})},\ \Eprint
  {http://arxiv.org/abs/1102.1014} {arXiv:1102.1014 [astro-ph.CO]} \BibitemShut
  {NoStop}%
%\%CITATION = ARXIV:1102.1014;\%\%
\bibitem [{\citenamefont {Tojeiro}\ \emph {et~al.}(2012)\citenamefont
  {Tojeiro}, \citenamefont {Percival}, \citenamefont {Brinkmann}, \citenamefont
  {Brownstein}, \citenamefont {Eisenstein} \emph {et~al.}}]{Tojeiro:2012rp}%
  \BibitemOpen
  \bibfield  {author} {\bibinfo {author} {\bibfnamefont {Rita}\ \bibnamefont
  {Tojeiro}}, \bibinfo {author} {\bibfnamefont {W.J.}\ \bibnamefont
  {Percival}}, \bibinfo {author} {\bibfnamefont {J.}~\bibnamefont {Brinkmann}},
  \bibinfo {author} {\bibfnamefont {J.R.}\ \bibnamefont {Brownstein}}, \bibinfo
  {author} {\bibfnamefont {D.}~\bibnamefont {Eisenstein}},  \emph {et~al.},\
  }\bibfield  {title} {\enquote {\bibinfo {title} {{The clustering of galaxies
  in the SDSS-III Baryon Oscillation Spectroscopic Survey: measuring structure
  growth using passive galaxies}},}\ }\href@noop {} {\  (\bibinfo {year}
  {2012})},\ \Eprint {http://arxiv.org/abs/1203.6565} {arXiv:1203.6565
  [astro-ph.CO]} \BibitemShut {NoStop}%
%\%CITATION = ARXIV:1203.6565;\%\%
\bibitem [{\citenamefont {Blake}\ \emph {et~al.}(2012)\citenamefont {Blake},
  \citenamefont {Brough}, \citenamefont {Colless}, \citenamefont {Contreras},
  \citenamefont {Couch} \emph {et~al.}}]{Blake:2012pj}%
  \BibitemOpen
  \bibfield  {author} {\bibinfo {author} {\bibfnamefont {Chris}\ \bibnamefont
  {Blake}}, \bibinfo {author} {\bibfnamefont {Sarah}\ \bibnamefont {Brough}},
  \bibinfo {author} {\bibfnamefont {Matthew}\ \bibnamefont {Colless}}, \bibinfo
  {author} {\bibfnamefont {Carlos}\ \bibnamefont {Contreras}}, \bibinfo
  {author} {\bibfnamefont {Warrick}\ \bibnamefont {Couch}},  \emph {et~al.},\
  }\bibfield  {title} {\enquote {\bibinfo {title} {{The WiggleZ Dark Energy
  Survey: Joint measurements of the expansion and growth history at z 1}},}\
  }\href {\doibase 10.1111/j.1365-2966.2012.21473.x} {\bibfield  {journal}
  {\bibinfo  {journal} {Mon.Not.Roy.Astron.Soc.}\ }\textbf {\bibinfo {volume}
  {425}},\ \bibinfo {pages} {405--414} (\bibinfo {year} {2012})},\ \Eprint
  {http://arxiv.org/abs/1204.3674} {arXiv:1204.3674 [astro-ph.CO]} \BibitemShut
  {NoStop}%
%\%CITATION = ARXIV:1204.3674;\%\%
\bibitem [{\citenamefont {de~la Torre}\ \emph {et~al.}(2013)\citenamefont
  {de~la Torre}, \citenamefont {Guzzo}, \citenamefont {Peacock}, \citenamefont
  {Branchini}, \citenamefont {Iovino} \emph {et~al.}}]{delaTorre:2013rpa}%
  \BibitemOpen
  \bibfield  {author} {\bibinfo {author} {\bibfnamefont {S.}~\bibnamefont
  {de~la Torre}}, \bibinfo {author} {\bibfnamefont {L.}~\bibnamefont {Guzzo}},
  \bibinfo {author} {\bibfnamefont {J.A.}\ \bibnamefont {Peacock}}, \bibinfo
  {author} {\bibfnamefont {E.}~\bibnamefont {Branchini}}, \bibinfo {author}
  {\bibfnamefont {A.}~\bibnamefont {Iovino}},  \emph {et~al.},\ }\bibfield
  {title} {\enquote {\bibinfo {title} {{The VIMOS Public Extragalactic Redshift
  Survey (VIPERS). Galaxy clustering and redshift-space distortions at z=0.8 in
  the first data release}},}\ }\href@noop {} {\  (\bibinfo {year} {2013})},\
  \Eprint {http://arxiv.org/abs/1303.2622} {arXiv:1303.2622 [astro-ph.CO]}
  \BibitemShut {NoStop}%
%\%CITATION = ARXIV:1303.2622;\%\%
\bibitem [{\citenamefont {Suzuki}\ \emph {et~al.}(2012)\citenamefont {Suzuki},
  \citenamefont {Rubin}, \citenamefont {Lidman}, \citenamefont {Aldering},
  \citenamefont {Amanullah} \emph {et~al.}}]{Suzuki:2011hu}%
  \BibitemOpen
  \bibfield  {author} {\bibinfo {author} {\bibfnamefont {N.}~\bibnamefont
  {Suzuki}}, \bibinfo {author} {\bibfnamefont {D.}~\bibnamefont {Rubin}},
  \bibinfo {author} {\bibfnamefont {C.}~\bibnamefont {Lidman}}, \bibinfo
  {author} {\bibfnamefont {G.}~\bibnamefont {Aldering}}, \bibinfo {author}
  {\bibfnamefont {R.}~\bibnamefont {Amanullah}},  \emph {et~al.},\ }\bibfield
  {title} {\enquote {\bibinfo {title} {{The Hubble Space Telescope Cluster
  Supernova Survey: V. Improving the Dark Energy Constraints Above z=1 and
  Building an Early-Type-Hosted Supernova Sample}},}\ }\href@noop {} {\bibfield
   {journal} {\bibinfo  {journal} {Astrophys.J.}\ }\textbf {\bibinfo {volume}
  {746}},\ \bibinfo {pages} {85} (\bibinfo {year} {2012})},\ \Eprint
  {http://arxiv.org/abs/1105.3470} {arXiv:1105.3470 [astro-ph.CO]} \BibitemShut
  {NoStop}%
%\%CITATION = ARXIV:1208.4855;\%\%
\bibitem [{\citenamefont {{Chevallier}}\ and\ \citenamefont
  {{Polarski}}(2001)}]{2001IJMPD..10..213C}%
  \BibitemOpen
  \bibfield  {author} {\bibinfo {author} {\bibfnamefont {M.}~\bibnamefont
  {{Chevallier}}}\ and\ \bibinfo {author} {\bibfnamefont {D.}~\bibnamefont
  {{Polarski}}},\ }\bibfield  {title} {\enquote {\bibinfo {title}
  {{Accelerating Universes with Scaling Dark Matter}},}\ }\href {\doibase
  10.1142/S0218271801000822} {\bibfield  {journal} {\bibinfo  {journal}
  {International Journal of Modern Physics D}\ }\textbf {\bibinfo {volume}
  {10}},\ \bibinfo {pages} {213--223} (\bibinfo {year} {2001})},\ \Eprint
  {http://arxiv.org/abs/gr-qc/0009008} {gr-qc/0009008} \BibitemShut {NoStop}%
\bibitem [{\citenamefont {{Linder}}(2003)}]{2003PhRvL..90i1301L}%
  \BibitemOpen
  \bibfield  {author} {\bibinfo {author} {\bibfnamefont {E.~V.}\ \bibnamefont
  {{Linder}}},\ }\bibfield  {title} {\enquote {\bibinfo {title} {{Exploring the
  Expansion History of the Universe}},}\ }\href {\doibase
  10.1103/PhysRevLett.90.091301} {\bibfield  {journal} {\bibinfo  {journal}
  {Physical Review Letters}\ }\textbf {\bibinfo {volume} {90}},\ \bibinfo {eid}
  {091301} (\bibinfo {year} {2003})},\ \Eprint
  {http://arxiv.org/abs/astro-ph/0208512} {astro-ph/0208512} \BibitemShut
  {NoStop}%
\bibitem [{\citenamefont {Amendola}\ \emph {et~al.}(2014)\citenamefont
  {Amendola}, \citenamefont {Fogli}, \citenamefont {Guarnizo}, \citenamefont
  {Kunz},\ and\ \citenamefont {Vollmer}}]{Amendola:2013qna}%
  \BibitemOpen
  \bibfield  {author} {\bibinfo {author} {\bibfnamefont {Luca}\ \bibnamefont
  {Amendola}}, \bibinfo {author} {\bibfnamefont {Simone}\ \bibnamefont
  {Fogli}}, \bibinfo {author} {\bibfnamefont {Alejandro}\ \bibnamefont
  {Guarnizo}}, \bibinfo {author} {\bibfnamefont {Martin}\ \bibnamefont {Kunz}},
  \ and\ \bibinfo {author} {\bibfnamefont {Adrian}\ \bibnamefont {Vollmer}},\
  }\bibfield  {title} {\enquote {\bibinfo {title} {{Model-independent
  constraints on the cosmological anisotropic stress}},}\ }\href {\doibase
  10.1103/PhysRevD.89.063538} {\bibfield  {journal} {\bibinfo  {journal}
  {Phys.Rev.}\ }\textbf {\bibinfo {volume} {D89}},\ \bibinfo {pages} {063538}
  (\bibinfo {year} {2014})},\ \Eprint {http://arxiv.org/abs/1311.4765}
  {arXiv:1311.4765 [astro-ph.CO]} \BibitemShut {NoStop}%
%\%CITATION = ARXIV:1311.4765;\%\%
\bibitem [{\citenamefont {Akrami}\ \emph {et~al.}(2014)\citenamefont {Akrami},
  \citenamefont {Koivisto},\ and\ \citenamefont {Solomon}}]{Akrami:2014lja}%
  \BibitemOpen
  \bibfield  {author} {\bibinfo {author} {\bibfnamefont {Yashar}\ \bibnamefont
  {Akrami}}, \bibinfo {author} {\bibfnamefont {Tomi~S.}\ \bibnamefont
  {Koivisto}}, \ and\ \bibinfo {author} {\bibfnamefont {Adam~R.}\ \bibnamefont
  {Solomon}},\ }\bibfield  {title} {\enquote {\bibinfo {title} {{The nature of
  spacetime in bigravity: two metrics or none?}}}\ }\href@noop {} {\  (\bibinfo
  {year} {2014})},\ \Eprint {http://arxiv.org/abs/1404.0006} {arXiv:1404.0006
  [gr-qc]} \BibitemShut {NoStop}%
%\%CITATION = ARXIV:1404.0006;\%\%
\bibitem [{\citenamefont {de~Rham}\ \emph {et~al.}(2014)\citenamefont
  {de~Rham}, \citenamefont {Heisenberg},\ and\ \citenamefont
  {Ribeiro}}]{deRham:2014naa}%
  \BibitemOpen
  \bibfield  {author} {\bibinfo {author} {\bibfnamefont {Claudia}\ \bibnamefont
  {de~Rham}}, \bibinfo {author} {\bibfnamefont {Lavinia}\ \bibnamefont
  {Heisenberg}}, \ and\ \bibinfo {author} {\bibfnamefont {Raquel~H.}\
  \bibnamefont {Ribeiro}},\ }\bibfield  {title} {\enquote {\bibinfo {title}
  {{On couplings to matter in massive (bi-)gravity}},}\ }\href@noop {} {\
  (\bibinfo {year} {2014})},\ \Eprint {http://arxiv.org/abs/1408.1678}
  {arXiv:1408.1678 [hep-th]} \BibitemShut {NoStop}%
%%CITATION = ARXIV:1408.1678;%%
\bibitem [{\citenamefont {{Yamashita}}\ \emph {et~al.}(2014)\citenamefont
  {{Yamashita}}, \citenamefont {{De Felice}},\ and\ \citenamefont
  {{Tanaka}}}]{2014arXiv1408.0487Y}%
  \BibitemOpen
  \bibfield  {author} {\bibinfo {author} {\bibfnamefont {Y.}~\bibnamefont
  {{Yamashita}}}, \bibinfo {author} {\bibfnamefont {A.}~\bibnamefont {{De
  Felice}}}, \ and\ \bibinfo {author} {\bibfnamefont {T.}~\bibnamefont
  {{Tanaka}}},\ }\bibfield  {title} {\enquote {\bibinfo {title} {{Appearance of
  Boulware-Deser ghost in bigravity with doubly coupled matter}},}\ }\href@noop
  {} {\bibfield  {journal} {\bibinfo  {journal} {ArXiv e-prints}\ } (\bibinfo
  {year} {2014})},\ \Eprint {http://arxiv.org/abs/1408.0487} {arXiv:1408.0487
  [hep-th]} \BibitemShut {NoStop}%
\bibitem [{\citenamefont {{Noller}}\ and\ \citenamefont
  {{Melville}}(2014)}]{2014arXiv1408.5131N}%
  \BibitemOpen
  \bibfield  {author} {\bibinfo {author} {\bibfnamefont {J.}~\bibnamefont
  {{Noller}}}\ and\ \bibinfo {author} {\bibfnamefont {S.}~\bibnamefont
  {{Melville}}},\ }\bibfield  {title} {\enquote {\bibinfo {title} {{The
  coupling to matter in Massive, Bi- and Multi-Gravity}},}\ }\href@noop {}
  {\bibfield  {journal} {\bibinfo  {journal} {ArXiv e-prints}\ } (\bibinfo
  {year} {2014})},\ \Eprint {http://arxiv.org/abs/1408.5131} {arXiv:1408.5131
  [hep-th]} \BibitemShut {NoStop}%
\bibitem [{\citenamefont {Enander}\ \emph {et~al.}(2014)\citenamefont
  {Enander}, \citenamefont {Solomon}, \citenamefont {Akrami},\ and\
  \citenamefont {M{\"o}rtsell}}]{Enander:2014xga}%
  \BibitemOpen
  \bibfield  {author} {\bibinfo {author} {\bibfnamefont {Jonas}\ \bibnamefont
  {Enander}}, \bibinfo {author} {\bibfnamefont {Adam~R.}\ \bibnamefont
  {Solomon}}, \bibinfo {author} {\bibfnamefont {Yashar}\ \bibnamefont
  {Akrami}}, \ and\ \bibinfo {author} {\bibfnamefont {Edvard}\ \bibnamefont
  {M{\"o}rtsell}},\ }\bibfield  {title} {\enquote {\bibinfo {title} {{Cosmic
  expansion histories in massive bigravity with symmetric matter coupling}},}\
  }\href@noop {} {\  (\bibinfo {year} {2014})},\ \Eprint
  {http://arxiv.org/abs/1409.2860} {arXiv:1409.2860 [astro-ph.CO]} \BibitemShut
  {NoStop}%
%%CITATION = ARXIV:1409.2860;%%
\bibitem [{\citenamefont {{Solomon}}\ \emph
  {et~al.}(2014{\natexlab{b}})\citenamefont {{Solomon}}, \citenamefont
  {{Enander}}, \citenamefont {{Akrami}}, \citenamefont {{Koivisto}},
  \citenamefont {{K{\"o}nnig}},\ and\ \citenamefont
  {{M{\"o}rtsell}}}]{2014arXiv1409.8300S}%
  \BibitemOpen
  \bibfield  {author} {\bibinfo {author} {\bibfnamefont {A.~R.}\ \bibnamefont
  {{Solomon}}}, \bibinfo {author} {\bibfnamefont {J.}~\bibnamefont
  {{Enander}}}, \bibinfo {author} {\bibfnamefont {Y.}~\bibnamefont {{Akrami}}},
  \bibinfo {author} {\bibfnamefont {T.~S.}\ \bibnamefont {{Koivisto}}},
  \bibinfo {author} {\bibfnamefont {F.}~\bibnamefont {{K{\"o}nnig}}}, \ and\
  \bibinfo {author} {\bibfnamefont {E.}~\bibnamefont {{M{\"o}rtsell}}},\
  }\bibfield  {title} {\enquote {\bibinfo {title} {{Does massive gravity have
  viable cosmologies?}}}\ }\href@noop {} {\bibfield  {journal} {\bibinfo
  {journal} {ArXiv e-prints}\ } (\bibinfo {year} {2014}{\natexlab{b}})},\
  \Eprint {http://arxiv.org/abs/1409.8300} {arXiv:1409.8300} \BibitemShut
  {NoStop}%
\end{thebibliography}%

\end{document}